\documentclass[fleqn,usenatbib]{mnras}

\usepackage{newtxtext,newtxmath}

\usepackage[T1]{fontenc}

\DeclareRobustCommand{\VAN}[3]{#2}
\let\VANthebibliography\thebibliography
\def\thebibliography{\DeclareRobustCommand{\VAN}[3]{##3}\VANthebibliography}

\usepackage{graphicx}	
\usepackage{amsmath}	

\title[Short title, max. 45 characters]{Collisional excitation of methyl (iso)cyanide by He atoms: rate coefficients and isomerism effects}

\author[M. Ben Khalifa et al.]{
M. Ben Khalifa$^{1}$\thanks{E-mail: malek.benkhalifa@kuleuven.be}
P. J. Dagdigian $^{2}$ and J. Loreau $^{1}$\thanks{E-mail: jerome.loreau@kuleuven.be}
\\
$^{1}$KU Leuven, Department of Chemistry, Celestijnenlaan 200F, B-3001 Leuven, Belgium.\\
$^{2}$Department of Chemistry, The Johns Hopkins University, Baltimore, MD 21218-2685, USA.\\
}

\date{Accepted XXX. Received YYY; in original form ZZZ}

\pubyear{2023}

\begin{document}
\label{firstpage}
\pagerange{\pageref{firstpage}--\pageref{lastpage}}
\maketitle

\begin{abstract}
Among all closed-shell species observed in molecular clouds, molecules with C$_{3v}$ symmetry play a crucial role, as their rotational spectroscopy allows them to behave as a gas thermometer.
In the interstellar medium, methyl cyanide (CH$_3$CN) is the second most abundant of those (after ammonia, NH$_3$). Its isomer, methyl isocyanide (CH$_3$NC) is less abundant but has been detected in many astrophysical sources. In order to assess their absolute and relative
abundances, it is essential to understand their collisional excitation properties. This paper reports the calculation of rate coefficients for rotational excitation of CH$_3$CN and CH$_3$NC molecules with He atoms, from low (5 K) to moderate (100 K) temperatures. We include the first 74 and 66 rotational states of both $para$ and $ortho$ symmetries of
CH$_3$CN and CH$_3$NC, respectively. A propensity for $\Delta j=2$ transitions is observed in the case of CH$_3$CN-He collisions, whereas in the case of CH$_3$NC-He a propensity for $\Delta j=1$ is observed for transitions involving low values of $j$ and at low temperatures, while a propensity for $\Delta j=2$ is observed for higher values of $j$ and at high temperatures.
A comparison of rate coefficients shows differences up to a factor of 3, depending on temperature and on the $ortho$/$para$ symmetries for dominant transitions. This confirms the importance of having specific collisional data for each isomer.
We also examined the effect of these new rates on the CH$_3$CN and CH$_3$NC excitation in molecular clouds by performing radiative transfer calculations
of the excitation and brightness temperatures for several detected lines.  

\end{abstract}




\section{Introduction}
Among all molecules detected in astrophysical clouds, nitriles (or cyanides, R-C$\equiv$N) are the largest class of compounds. They represent about one-fifth of the more than 250 molecules
identified to date in the interstellar medium (ISM) and circumstellar shells \citep{muller2001cologne,mcguire20222021}. Cyanides are readily detected in a number of interstellar environments thanks to their
large dipole moments \citep{lee1972microwave}. Recently, even 
complex cyanides like hydroxyacetonitrile (HOCH$_2$C$\equiv$N) \citep{zeng2019first} and the aromatic ring benzonitrile (C$_6$H$_5$CN) \citep{mcguire2018detection} have been detected. 
In addition to nitrile molecules, a few isocyanides (with the functional
group -N$\equiv$C) have also been detected in objects of the ISM. They include HNC \citep{bellili2019single,schilke2001line}, HNCO \citep{snyder1972detection}, CH$_3$NC \citep{irvine1984cyanide,cernicharo1988tentative,remijan2005interstellar,margules2018submillimeter,calcutt2018alma} HCCNC \citep{kawaguchi1992detection}, CH$_2$CHNC (tentatively) \citep{lopez2014laboratory} as well as inorganic isocyanides \citet{muller2001cologne}.

CH$_3$CN is one of the most ubiquitous interstellar organic molecules and one that is regularly  detected. It was observed in a variety of
low- and high-mass sources \citep{beltran2005detailed,zapata2010rotating}. 
CH$_3$CN was detected towards the massive star forming regions Sagittarius A and B, close to the Galactic center \citep{pols2018physical}. It has also been identified ubiquitously, e.g. in the TMC-1 dark cloud \citep{matthews1983detection}, the low-mass protostar IRAS 16293-2422 \citep{cazaux2003hot}, in the circumstellar shell of the carbon-rich star IRC +10216 \citep{agundez2008detection}, and in comets such as Kohoutek \citep{ulich1974detection}. Recently, large quantities of CH$_3$CN were detected in the protoplanetary disk surrounding the young star MWC-480 \citep{oberg2015comet}.
Thanks to its three-fold symmetry and its large dipole moment, CH$_3$CN is used as a gas thermometer to probe interstellar environments and it is a useful tracer of the dense gas \citep{wilner1994maps}. 
Its isomer CH$_3$NC is thermodynamically the less stable isomer of CH$_3$CN by about 103 kJ/mol, with a barrier to isomerization of about 160 kJ/mol \citep{nguyen2018unimolecular}. 
CH$_3$NC was tentatively observed for the first time in the Sgr B2 cloud \citet{cernicharo1988tentative} and confirmed with additional transitions seventeen years later by \citet{remijan2005interstellar}.
Since then, it has been observed only in a handful of sources, including the Horsehead nebula photodissociation region by \citet{gratier2013iram} and Orion KL by \citet{lopez2014laboratory}, among others \citep{margules2018submillimeter,calcutt2018alma}.

For typical dense cloud conditions, theoretical calculations of \citet{defrees1985theoretical} estimated the ratio of CH$_3$NC/CH$_3$CN to be in the range of 0.1-0.4. However, \citet{cernicharo1988tentative} deduced from their
observations an abundance ratio of $\sim$ 0.03-0.05, showing that CH$_3$CN is an order of magnitude higher than previous estimates. Later, \citet{gratier2013iram} found, according to their observations in the Horsehead region,
an abundance ratio CH$_3$NC/CH$_3$CN in agreement with estimations of \citet{defrees1985theoretical}.

In astrophysical media, the chemistry of CH$_3$CN is assumed to be governed by two processes: the dissociative recombination reaction CH$_3$CNH$^+$ + e$^-$ $ \rightarrow$ CH$_3$CN + H, 
 and by the radiative association mechanism CN + CH$_3$  
 $\rightarrow$ CH$_3$CN + $h\nu$ \citep{mcelroy2013umist,mackay1999ch3cn,willacy1993gas}, whereas gas-grain chemistry can also provide an important pathway for the formation of methyl cyanide, either from direct addition of CH$_3$ and CN or from successive hydrogenations starting with C$_2$N \citep{belloche2009increased}.
Acetonitrile and methyl isocyanide are isoelectronic symmetric top molecules which show close similarities in their molecular parameters such as the bond length \citep{moffat1964lcao,costain1958determination}
and electric dipole moment \citep{lee1972microwave}, which cannot introduce any bias in their respective rotational spectra. Therefore, the formation and evolution of interstellar clouds and the stars they generate
can be explored through isomer comparison studies.

In order to identify molecules in the ISM and their densities, a precise determination of the positions and intensity of spectral lines is necessary. Although the vibration-rotation spectrum of  both CH$_3$CN and CH$_3$NC is well established, their collisional properties are not well constrained, which can lead to errors on the inferred densities.

In the ISM, the local thermodynamic equilibrium (LTE) conditions are usually not fulfilled \citep{roueff2013molecular}. Therefore, both collisional and radiative processes should be considered in the molecular line modeling in order to derive relative abundances or column densities of CH$_3$CN/NC as well as the physical conditions of the gas. Einstein coefficients corresponding to radiative processes are usually available in literature; however, collisional rates between CH$_3$CN/NC with the most abundant species (H$_2$, H and He) have to be computed.

To the best of our knowledge, the only rate coefficients for CH$_3$CN-He/H$_2$ that are available in literature have been computed by \citet{green1986collisional}. These collisional rate coefficients are based on the Infinite Order Sudden (IOS) approximation, without any potential calculation, which can lead to large errors. In addition, the rates published by Green begin at a temperature of 20 K, preventing a precise modelling in the dense molecular clouds, with a typical temperature around 5 K. Moreover, concerning the isomer CH$_3$NC, no collision rate coefficients seem to be available and the interpretation of astrophysical observation is always realized by using rate coefficients of CH$_3$CN. 

For a precise determination of the absolute and relative abundance of methyl (iso)cyanide, we present in this paper the first rate coefficients for the collision excitation of CH$_3$NC, which we compare to those obtained for CH$_3$CN to highlight isomerism effects.
We restrict ourselves to the collisional excitation of the rigid symmetric top molecules CH$_3$CN and its isomer CH$_3$NC with He atoms, which are investigated with the same computational method.

The overall structure of this paper is as follows: first, we present in section~\ref{PES}, a brief description of the \textit{ab initio} calculations of CH$_3$CN/NC-He systems used in this work.
Section~\ref{Taux} describes the scattering calculations. We illustrate inelastic collisional rate coefficients of CH$_3$CN-He and CH$_3$NC-He complexes in section~\ref{Tauxx}. We analyze the
effect of our rate coefficients by performing a radiative transfer calculation for typical interstellar conditions in section~\ref{astro} .
Conclusions and future outlooks are drawn in Section~\ref{ccl}.

\section{Potential energy surfaces}\label{PES}
Recently, 3-dimensional potential energy surfaces (PESs) were obtained by \citet{ben2022interaction}, for the symmetric top molecules CH$_3$CN and CH$_3$NC in their ground electronic state ($\tilde{X}^1A_1$) interacting with He atoms. The PESs were computed using the explicitly correlated coupled cluster CCSD(T)-F12a method \citep{knizia2009simplified} in conjunction with the aug-cc-pVTZ basis set \citep{dunning1989gaussian} as implemented in the {\small MOLPRO} package \citep{werner2015molpro}.\\
The CH$_3$CN and CH$_3$NC molecules were considered as rigid rotors with the internal coordinates of methyl (iso)cyanide frozen at their experimental equilibrium values for the vibrational ground state.

To compute the interaction potentials, we chose the center of mass (c.m.) of
CH$_3$CN and CH$_3$NC molecules to be the origin of the coordinate system,
and we have characterized the orientations of each complex by $R$, $\theta$ and $\phi$, where $R$ represents the distance between the center of mass of CH$_3$CN/NC and the He atom, $\theta$ is the angle between \textbf{R} and the $C_3$ axis of the molecule, and $\phi$ is the azimuthal angle that describes the rotation of the He atom around the C$_{3v}$ axis. 

The global minimum for the CH$_3$CN-He PES is located at the configuration 
$\phi=60^{\circ}$,
$\theta=100^{\circ}$
and $R=$6.15 bohr with a well depth of 
55.10 cm$^{-1}$, while the global minimum of the CH$_3$NC-He PES has a well depth of 58.61 cm$^{-1}$ located at $\phi=60^{\circ}$, $\theta=100^{\circ}$ and $R=6.05$ bohr. These features are similar to those of PESs for other symmetric top + rare gas atom systems such as NH$_3$ with He, Ne, Ar, Kr, Xe
and SiH$_3$CN-He \citep{Loreau2014b,Loreau2015a,naouai2021inelastic}.
The potential energy surfaces of CH$_3$CN-He and CH$_3$NC-He were  fitted 
using the procedure given by \citet{green1976rotational} for the 
NH$_{3}$-He system. The dependence of the PESs on the He-CH$_3$CN and He-
CH$_3$NC angles was fitted by the usual spherical harmonic expansion (see Eq 1). From an \textit{ab initio} grid containing 19 $\theta$-values and 7 
$\phi$-values, we end up with 70 terms with $l_{max}$=18, with a
deviation between \textit{ab initio} potentials and analytical form less than 1\%. 

\begin{equation}\label{eq_expansion}
V(R,\theta,\phi)=\sum_{l=0}^{l_{max}}\sum_{m=0}^{l}V_{lm}(R)\frac{Y_{l}^{m}(\theta,\phi)+(-1)^{m}Y_{l}^{-m}(\theta,\phi)}{1+\delta_{m,0}}
\end{equation}
The shape and magnitude of the radial coefficients $V_{lm}$ reflect the stronger anisotropy of the interaction potential of CH$_3$NC-He compared to CH$_3$CN-He. While the global minimum of the PES is deeper for CH$_3$NC-He than for CH$_3$CN-He, the isotropic term is less deep for CH$_3$NC-He (56.2 cm$^{-1}$) than CH$_3$CN-He (65.8 cm$^{-1}$). Furthermore, for odd values of $l$, the radial coefficients are 
larger for CH$_3$NC-He than for CH$_3$CN-He, however, $V_{lm}$ with even 
values of $l$ are larger for CH$_3$CN than those for CH$_3$NC in the medium
and long-range of distances $R$, and smaller at short range. This is expected to have
an effect on the propensity rules in the rotational excitation, as shall be
further discussed in section~\ref{Tauxx}.

All technical details concerning the \textit{ab initio} PESs, potential well depths and their accurate locations as well as the fitting procedure, can be found in \citet{ben2022interaction}
(hereafter Paper I).

\section{Scattering calculations}\label{Taux}

\subsection{Cross sections}
Scattering calculations corresponding to rotational (de-)excitation of CH$_3$CN and CH$_3$NC molecules by collision with He atoms were performed for kinetic energies ranging from 0.1 to 900 cm$^{-1}$.

Methyl (iso)cyanide is a prolate symmetric top molecule. The rotational levels structure associated to both isomers were obtained using the $B$ and $A$
rotational constants fit on experimental spectra, which are equal to 0.3353 and 5.2420 cm$^{-1}$ for CH$_3$NC \cite{margules2001ab} and 0.3068 and 5.2470 cm$^{-1}$ for CH$_3$CN \citep{remijan2007alma}
and are function of two quantum numbers, $j$ and $k$, where $j$ is the rotational angular momentum and $k$ is a pseudo quantum number associated to the projection of $j$ on the body-fixed $z$ axis. Due to the centrifugal distortion, the rotational energy deviation between two adjacent $j$-levels decreases when $k$ increases. Therefore, it is possible to find multiple lines with various 
excitation energies in one spectrum, which makes CH$_3$CN a good probe of the local temperature.
The state of the nuclear spin associated to hydrogen atoms determines whether CH$_3$CN and CH$_3$NC have $para$ ($E$) or $ortho$ ($A$) symmetry.
 The spectral signatures are linked to $k$-ladder structures whose notation for the $ortho$ configuration is $k$=$3n$ and for the $para$ configuration is $k$=$3n\pm 1$, $n$ being an integer.  As the $ortho$ and $para$ levels of CH$_3$CN and CH$_3$NC cannot interchange in inelastic collisions, the state-to-state cross sections were computed separately for each nuclear spin species.

The most accurate method to compute state-to-state cross sections is the close-coupling (CC) approach. However, due to the small spacing between rotational levels and the fact that transitions involving highly excited rotational states have been observed (up to $j=18$), a large rotational basis set of CH$_3$CN and CH$_3$NC is needed, which makes the use of the full CC method \citep{arthurs1960theory} extremely expensive in computational time with  increasing energy.
In order to achieve a good compromise between the accuracy of the results and computational cost, the collisional cross sections were calculated using the CC method for $E_{\textrm{tot}} \leq $ 100 cm$^{-1}$, and with the coupled-state (CS) approximation \citep{mcguire1974quantum} for higher energies, up to $E_{\textrm{tot}}$=900 cm$^{-1}$ using both {\small MOLSCAT} \citep{hutson2019molscat2} and {\small HIBRIDON} \citep{hibridon2023} codes. We carried out convergence tests at different energies in order to guarantee the validity of CS approximation, as presented in Table~\ref{test}. As reported, the relative error between CC and CS approximation for several transitions does not exceed 10\% for $E_{\textrm{tot}}$= 100 cm$^{-1}$ and 5\% for $E_{\textrm{tot}}$= 300 cm$^{-1}$. Furthermore, CPU time and disk occupancy for CS computations are almost five times 
smaller than CC ones. Globally, the CS approximation reduced the computational cost without a significant loss
of accuracy, especially at high collision energy.

For astrophysical purposes, the highest rotational levels to be converged in our scattering calculations are $j_k$=16$_6$ for $ortho$-CH$_3$NC/CN (with $E_{\textrm{rot}}$= 267.84 cm$^{-1}$ and 262.07 cm$^{-1}$, respectively) and $j_k$=16$_5$ for $para$-CH$_3$NC/CN (with $E_{\textrm{rot}}$= 213.87 cm$^{-1}$ and 207.52 cm$^{-1}$), respectively. To do so, the rotational basis set has been extended enough to ensure the convergence of the collisional cross-sections. 

The number of rotational 
levels $N$ considered for our computations was taken as : $N=35$ (up to $j_k$=20$_0$ with an energy of 140.8 cm$^{-1}$) for total energies E$_{\textrm{tot}} \leq$ 50 cm$^{-1}$, $N=55$ ( up to $j_k$=26$_0$ with an energy of 235.4 cm$^{-1}$) for total energies 50 cm$^{-1}$ $\leq$ E$_{\textrm{tot}} \leq$ 100 cm$^{-1}$, and $N=120$ ( up to $j_k$=29$_9$ with an energy of 689.17 cm$^{-1}$) for 100 cm$^{-1}$ $\leq$ E$_{\textrm{tot}} \leq$ 300 cm$^{-1}$ for $ortho$-CH$_3$NC, while for $ortho$-CH$_3$CN it was necessary to select 
$N=55$ (all levels up to $j_k$= 12$_6$, energy of 225.6 cm$^{-1}$) for total energies E$_{\textrm{tot}} \leq$ 50 cm$^{-1}$, 65 (up to $j_k$= 29$_3$ energy of 311.4 cm$^{-1}$) for 50 cm$^{-1}$ $\leq$ E$_{\textrm{tot}} \leq$ 100 cm$^{-1}$, and 115 ( up to $j_k$=35$_6$ with an energy of 564.33 cm$^{-1}$) for 100 cm$^{-1}$ $\leq$ E$_{\textrm{tot}} \leq$ 300 cm$^{-1}$. At the largest total energy considered (900 cm$^{-1}$), the rotational basis were extended to $N= 160$ or $N= 150$, for CH$_3$NC and CH$_3$CN, respectively.

For $para$-CH$_3$NC-He, we take $N= 65$ (up to $j_k$= 13$_5$, with an energy of 183.7 cm$^{-1}$)
for $E_{\textrm{tot}} \leq$ 50 cm$^{-1}$, $N=75$ (up to $j_k$= 20$_4$ with an energy of 219.3 cm$^{-1}$) for 50 cm$^{-1}$ $\leq$ E$_{\textrm{tot}} \leq$ 100 cm$^{-1}$, $N=145$ (up to $j_k$= 17$_8$ with an energy of 416.63 cm$^{-1}$) for 100 $\leq$
E$_{\textrm{tot}} \leq$ 300 and $N=180$ {\color{black}(up to $j_k=14_{10}$ with an energy of 561.085 cm$^{-1}$)} for 300 $\leq$ E$_{\textrm{tot}} \leq$ 900 cm$^{-1}$, while for $para$-CH$_3$CN, we take
$N= 78$ for E$_{\textrm{tot}} \leq$ 100 cm$^{-1}$ (up to $j_k$= 18$_7$, with an energy 346.9 cm$^{-1}$) and $N= 202$ for 100 cm$^{-1}$ $\leq$ $E_{\textrm{tot}}$ $\leq$ 900 cm$^{-1}$ (up to $j_k$= 14$_{13}$, with an energy 898.8 cm$^{-1}$).
The maximum value of the total angular momentum $J$ was selected such that the inelastic cross sections are all converged to within 0.05~\AA$^2$.
Finally, the integral cross sections were obtaining by summing all partial cross sections over the total angular momentum $J$ until reaching the convergence.

We computed collisional cross sections for total energy ranging from 0.1 to 900 cm$^{-1}$. In order to describe the resonances in the cross sections, we used a smaller energy step $dE$ at low collision energy and a larger energy step as the collisional energy increases and the resonances vanish. 
We used an energy step $dE$ of 0.2 cm$^{-1}$ for energies below 50 cm$^{-1}$, 0.5 cm$^{-1}$ for energies between 50 cm$^{-1}$ and 100 cm$^{-1}$, 2 cm$^{-1}$ for energies between 100 cm$^{-1}$ and 200 cm$^{-1}$,  5 cm$^{-1}$ for energies between 200 cm$^{-1}$ and 400 cm$^{-1}$, and  10 cm$^{-1}$ for energies between 400 cm$^{-1}$ and 900 cm$^{-1}$.

\begin{table}
\centering
\caption{Comparison between CC and CS cross sections (in ~\AA$^2$) for the excitation of $para$ and $ortho$ CH$_3$NC by He for total energies $E_{\textrm{tot}}$=100 and 300 cm$^{-1}$ (For $E_{\textrm{tot}}$=300 cm$^{-1}$, total angular momenta $J$ up to 20 are considered).}\label{test}
\begin{tabular}{ccccc}
\hline
\hline
Molecule & Energy & $j_k \rightarrow j'_k$ & CC & CS\\
\hline
o-CH$_3$NC & $E$=100 cm$^{-1}$ & 0$_0 \rightarrow 1_0$ & 7.208 &  7.042 \\
  &                & 0$_0 \rightarrow 2_0$ & 14.36 &  14.11 \\
  &                & 0$_0 \rightarrow 3_0$ & 2.742 &  2.944 \\
  &                & 0$_0 \rightarrow 4_0$ & 8.407 &  8.016 \\
\hline
o-CH$_3$NC & $E$=300 cm$^{-1}$ & 0$_0 \rightarrow 1_0$ & 0.500 &  0.498 \\
             &     & 0$_0 \rightarrow 2_0$ & 1.089 &  0.981 \\
             &     & 0$_0 \rightarrow 3_0$ & 0.307 &  0.290 \\
             &     & 0$_0 \rightarrow 4_0$ & 0.983 &  0.950 \\ 
\hline
p-CH$_3$NC & $E$=100 cm$^{-1}$ & 1$_1 \rightarrow 2_1$ & 8.342 &  8.628 \\
             &     & 1$_1 \rightarrow 3_1$ & 9.847 &  10.13 \\
             &    & 1$_1 \rightarrow 4_1$ & 2.887 &  3.091 \\
             &     & 1$_1 \rightarrow 5_1$ & 7.887 &  7.438 \\ 
\hline
p-CH$_3$NC & $E$=300 cm$^{-1}$ & 1$_1 \rightarrow 2_1$ & 0.475 &  0.465 \\
             &     & 1$_1 \rightarrow 3_1$ & 0.535 &  0.526 \\
             &     & 1$_1 \rightarrow 4_1$ & 0.585 &  0.571 \\
             &     & 1$_1 \rightarrow 5_1$ & 0.527 &  0.420 \\ 
\hline
\end{tabular}
\end{table}

\subsection{Rate coefficients}\label{Tauxx}
From the collisional cross-sections $\sigma_{i \rightarrow f}(E_c)$, we derive the corresponding collisional rate coefficients $k_{if}$ at a temperature $T$ by an average over a Maxwellian distribution of kinetic energy as expressed:
\begin{equation}
 k_{i \rightarrow f}(T)=\biggl(\frac{8}{\pi\mu\beta}\biggl)^{\frac{1}{2}}\beta^2\int_0^{\infty} E_c \sigma_{i \rightarrow f}(E_c)e^{-\beta E_c} dE_c
\end{equation}
Where $\beta$=$1/k_BT$, and $k_B$, $T$ and $\mu=3.646815237$ a.u denote the Boltzmann constant, the kinetic temperature and the reduced mass of the system, respectively.

For CH$_3$CN-He and CH$_3$NC-He, the state-to-state cross sections computed for total energy up to 900 cm$^{-1}$ leads to converged collisional rate coefficients for the first 74 $para$ levels and 66 $ortho$ levels of each isomer up to a kinetic temperature $T=100$ K. 
This complete set of (de-)excitation rate coefficients will be made available online on the LAMDA, BASECOL and EMAA websites. 

\begin{figure*}
\centering
{\label{c}\includegraphics[width=.46\linewidth]{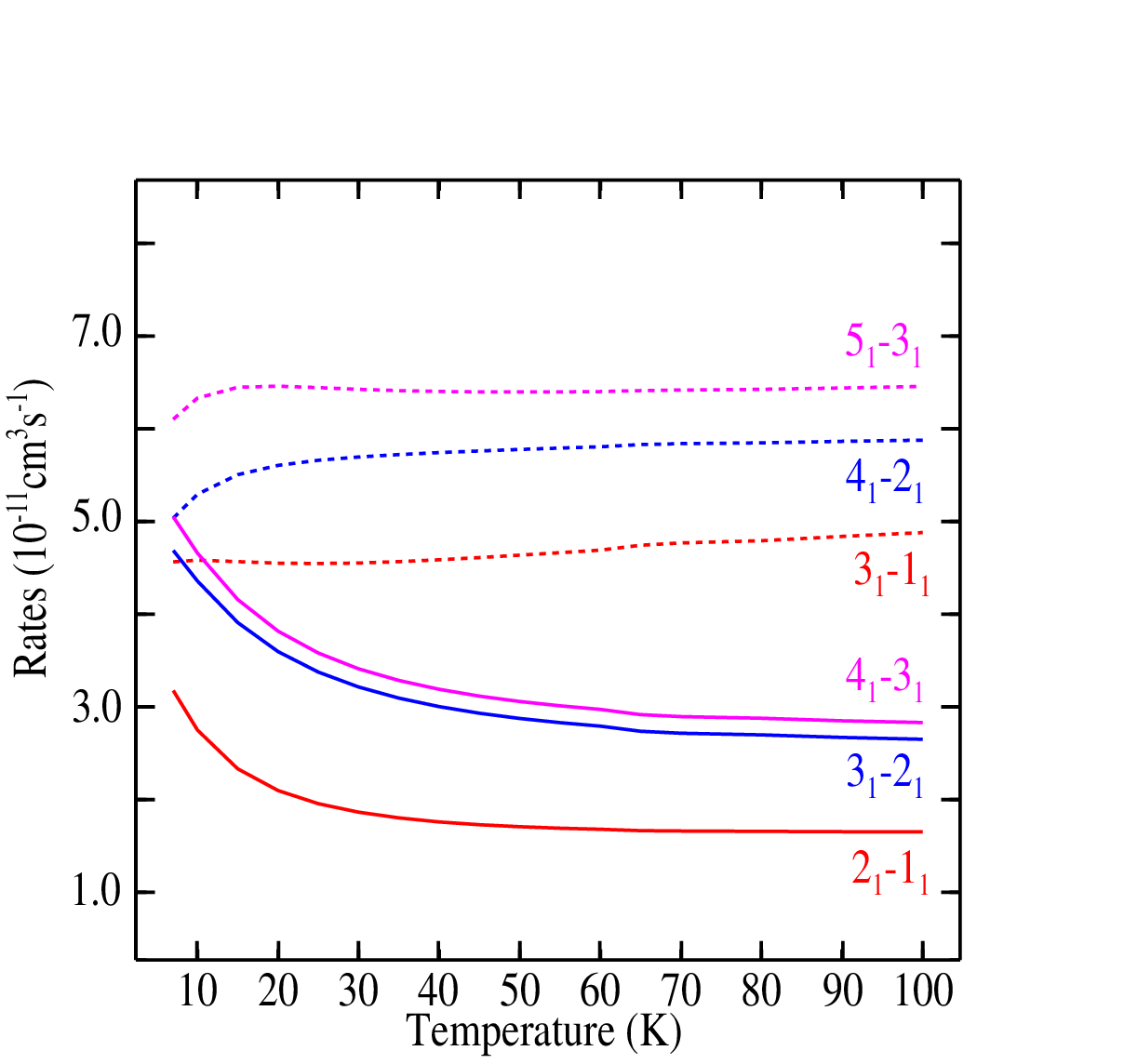}}
{\label{c}\includegraphics[width=.46\linewidth]{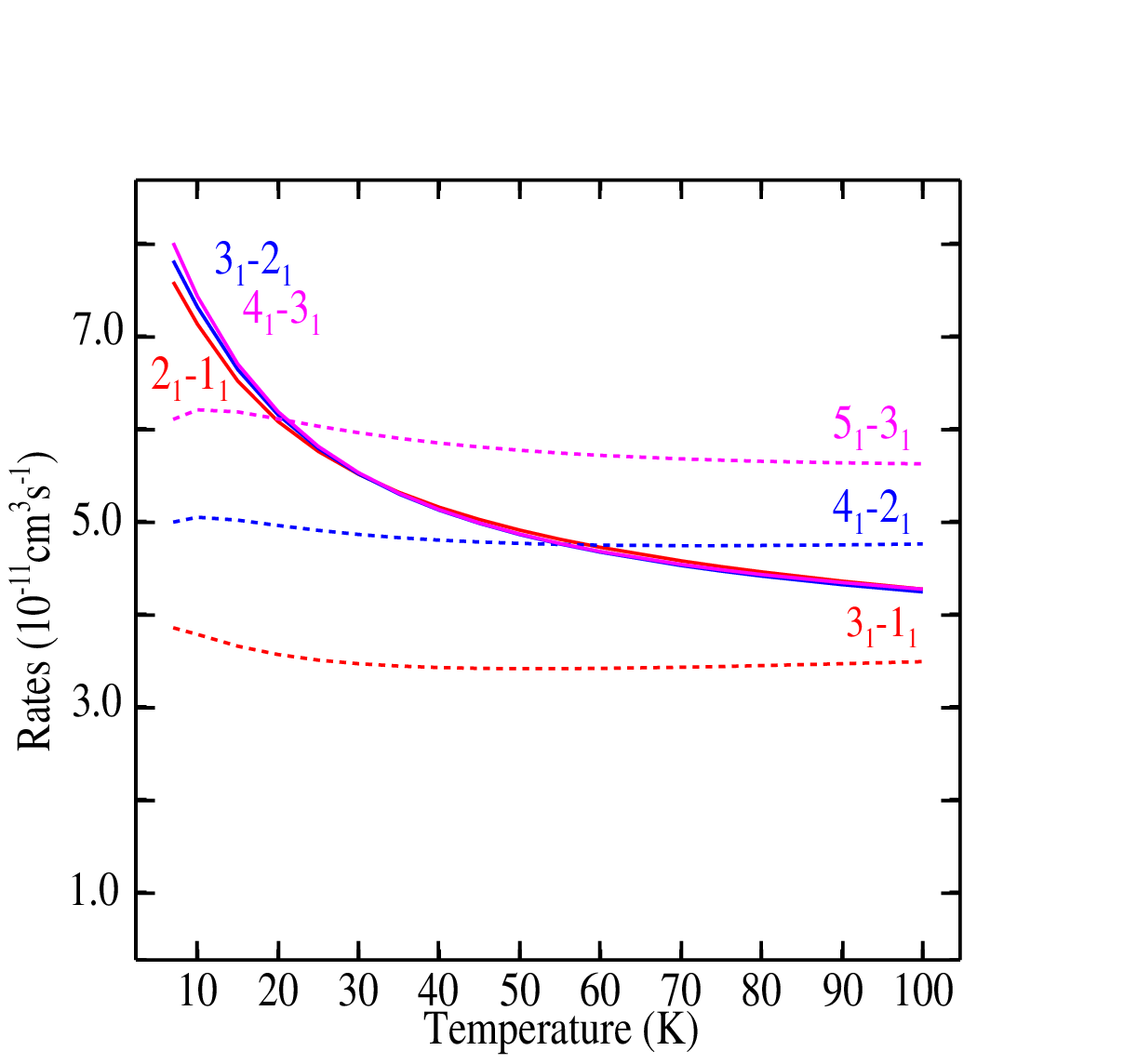}}
{\label{c}\includegraphics[width=.46\linewidth]{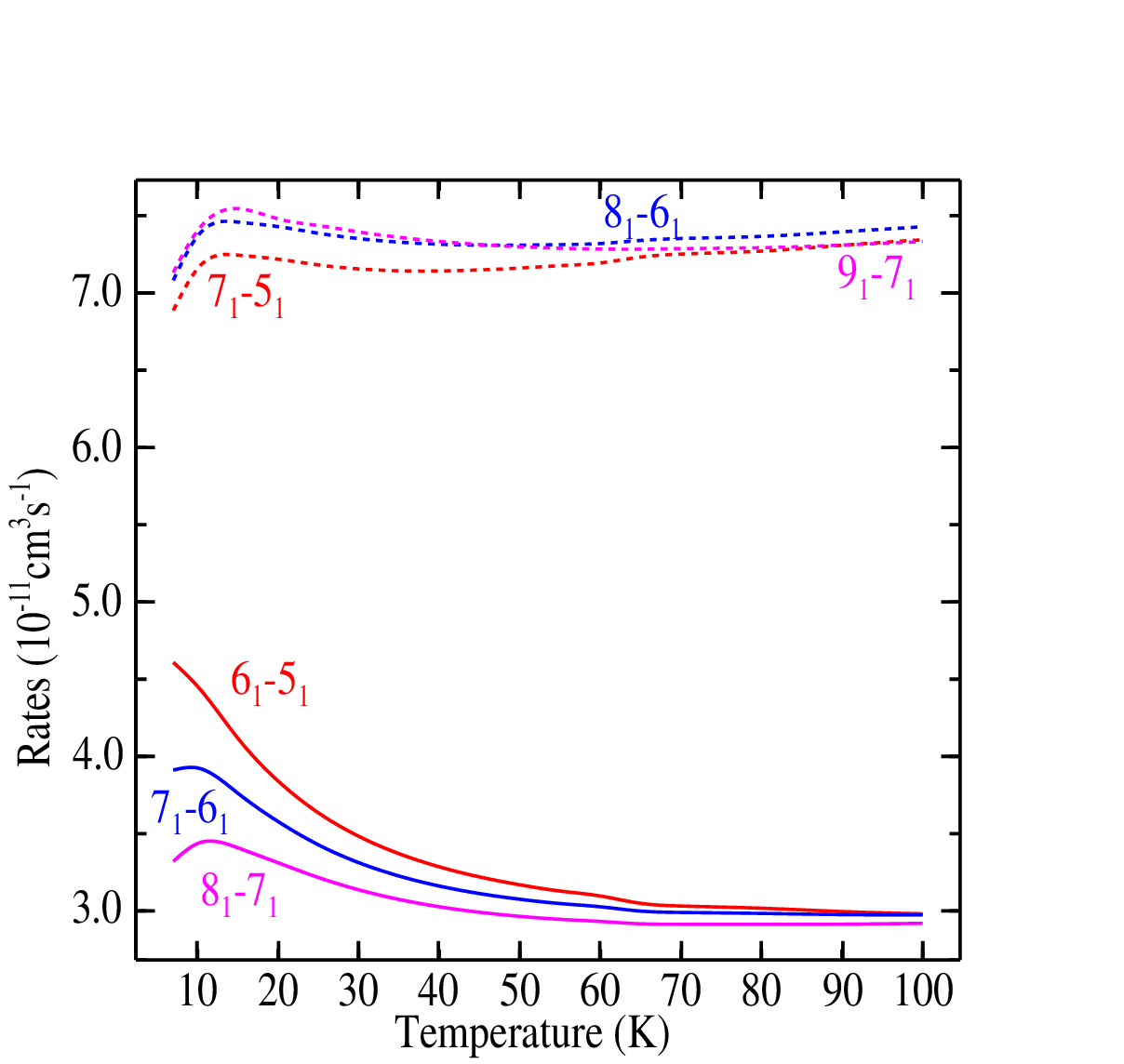}}
{\label{c}\includegraphics[width=.46\linewidth]{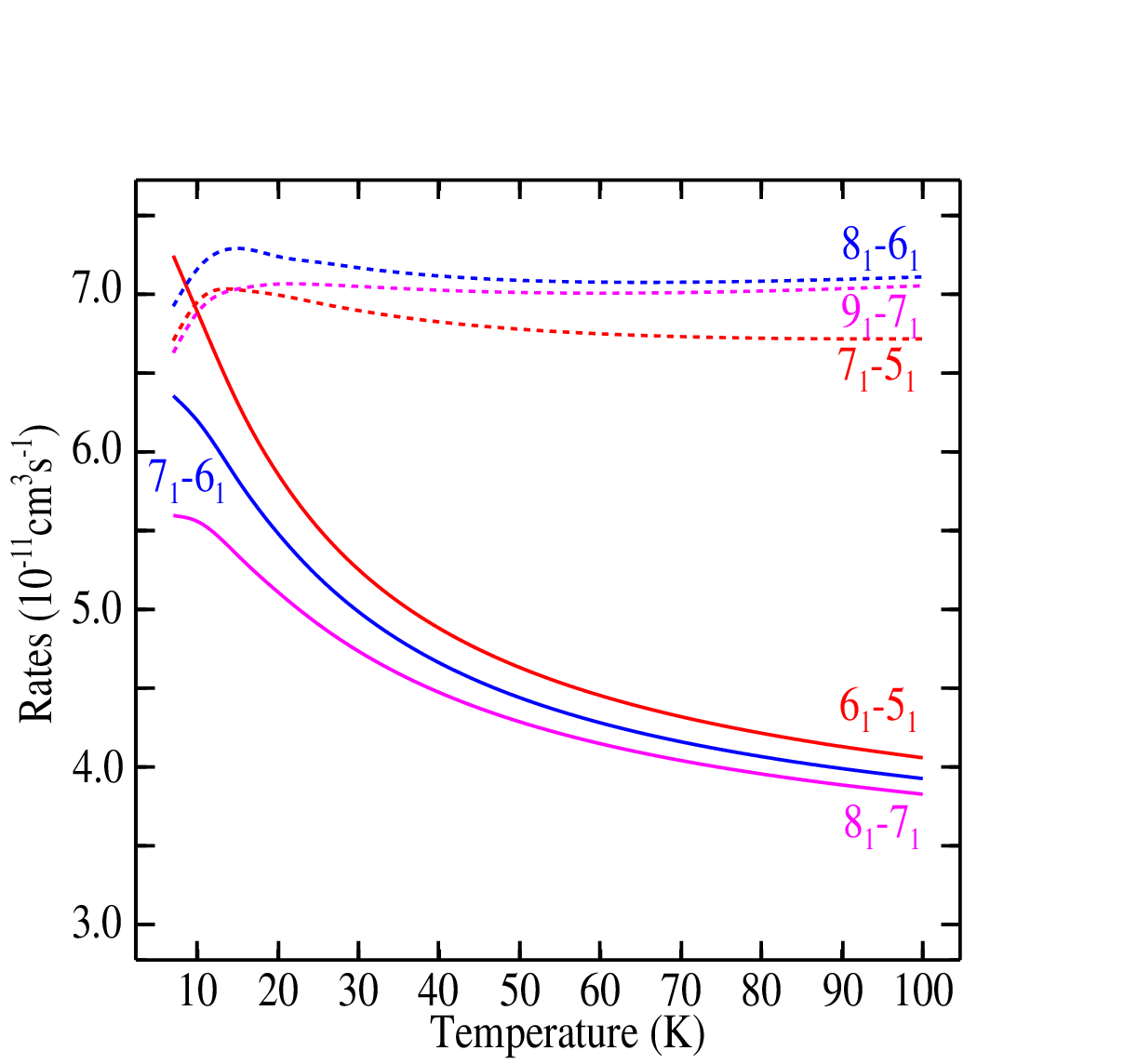}}
\caption{Temperature dependence of the rotational de-excitation rate coefficients for transitions $j_k \rightarrow j'_{k'}$ of $para-$CH$_3$CN-He (left panels) and $para-$CH$_3$NC-He (right panels) induced by collisions with He. Top panels: comparison of $\Delta j=1$ and $\Delta j=2$ transitions with $j^\prime=1-3$ and $k=k'=1$; Bottom panels: Same but  with $j^\prime=5-7$.}\label{Taux-E-CH3NC-CN-He}
\label{Taux-E-CH3NC-CN-He} 
\end{figure*}

We present in Fig.~\ref{Taux-E-CH3NC-CN-He}, the temperature variation of CH$_3$CN-He and CH$_3$NC-He state-to-state rate coefficients as a function of the temperature for selected transitions. We focus specifically on dipolar ($\Delta j$=1) and quadrupolar ($\Delta j$=2) transitions with $k=k^\prime=1$ in $para$-CH$_3$CN/NC. 
One can observe that there are significant differences between the CH$_3$CN-He and CH$_3$NC-He rate coefficients and that the propensity rules observed in the two sets of collisional rates are different. 

The rate coefficients for CH$_3$CN-He associated to $\Delta j =1$ transitions are considerably smaller than those for CH$_3$NC-He transitions, while those for $\Delta j =2$ are of similar magnitude. This difference can be understood on the basis of the expansion terms in Eq. \ref{eq_expansion}, and in particular the fact that the term $V_{10}$ that drives $\Delta j =1 $ transitions is much larger for CH$_3$NC-He than for CH$_3$CN-He, as already discussed in Paper I. Overall the PES of CH$_3$NC-He is deeper but it also has a larger degree of anisotropy than the PES of CH$_3$CN-He, leading to larger rate coefficients for $\Delta j =1 $ transitions.
For CH$_3$CN-He, the $\Delta j =2$ transitions dominate over $\Delta j =1$ transitions at all temperatures. 

On the other hand, for CH$_3$NC-He the dominant transitions are those associated with $\Delta j=1$ at low temperature, and $\Delta j=2$ at higher temperatures, leading to propensity rules that depend on the temperature. In addition, when the value of $j$ increases, the transitions $\Delta j=2$ become more and more dominant, including at low temperature. We can interpret this again based on the expansion of the PES. As discussed in Paper I, the $V_{10}$ term is the dominant term for CH$_3$NC-He at low energies (up to about 40 cm$^{-1}$), while $V_{20}$ dominates at higher energies. As a result, for highly excited rotational levels $V_{20}$ it always the dominant term in the expansion and $\Delta j=2$ transitions are favored.

While Fig.~\ref{Taux-E-CH3NC-CN-He} only illustrates rate coefficients for $para$-CH$_3$CN/NC-He and for transitions with $k=k^\prime=1$, the same behavior is also observed for other values of $k$ as well as for $ortho$-CH$_3$CN/NC-He.

\begin{figure*}
\centering
{\label{c}\includegraphics[width=.49\linewidth]{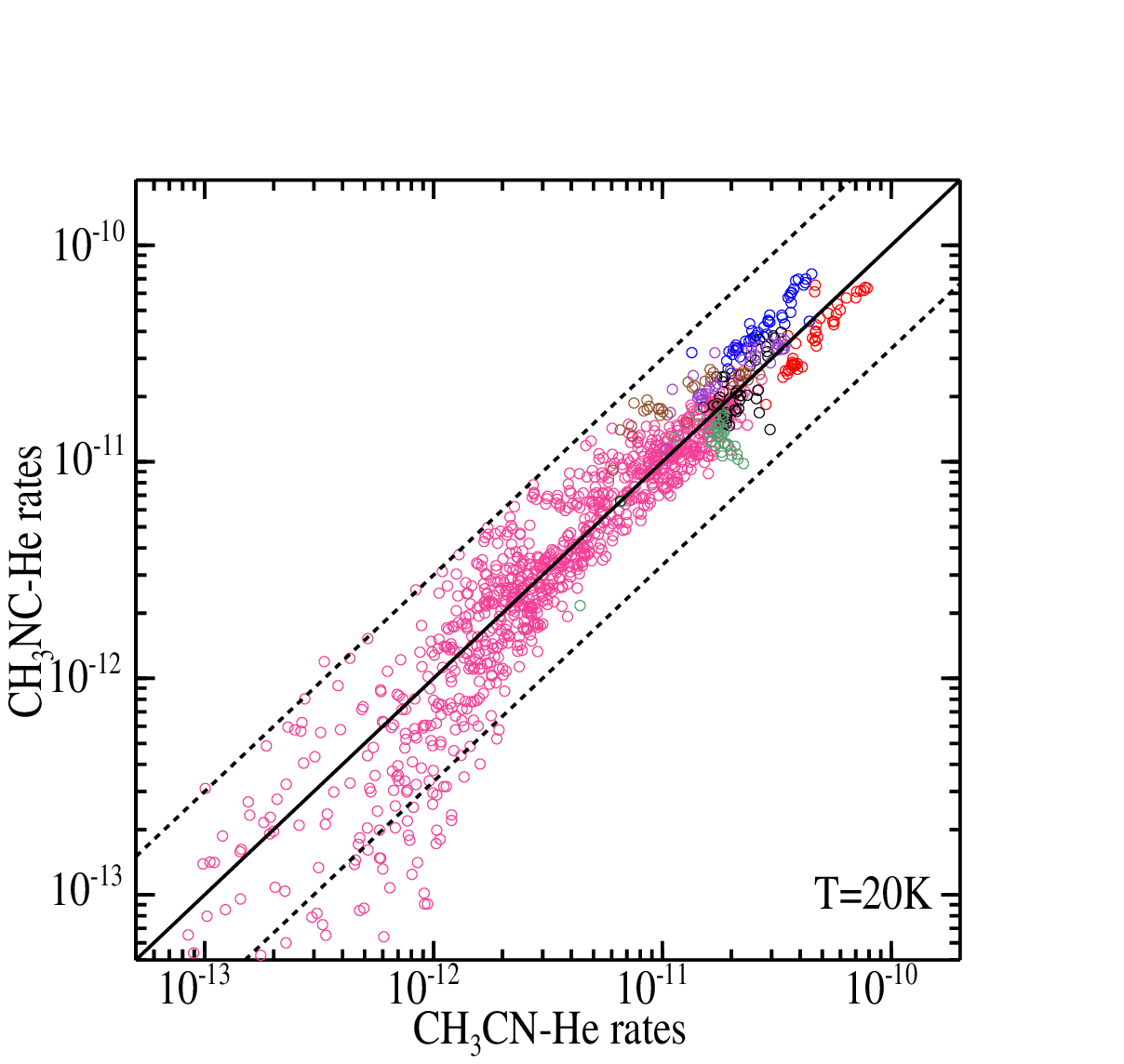}}
{\label{c}\includegraphics[width=.49\linewidth]{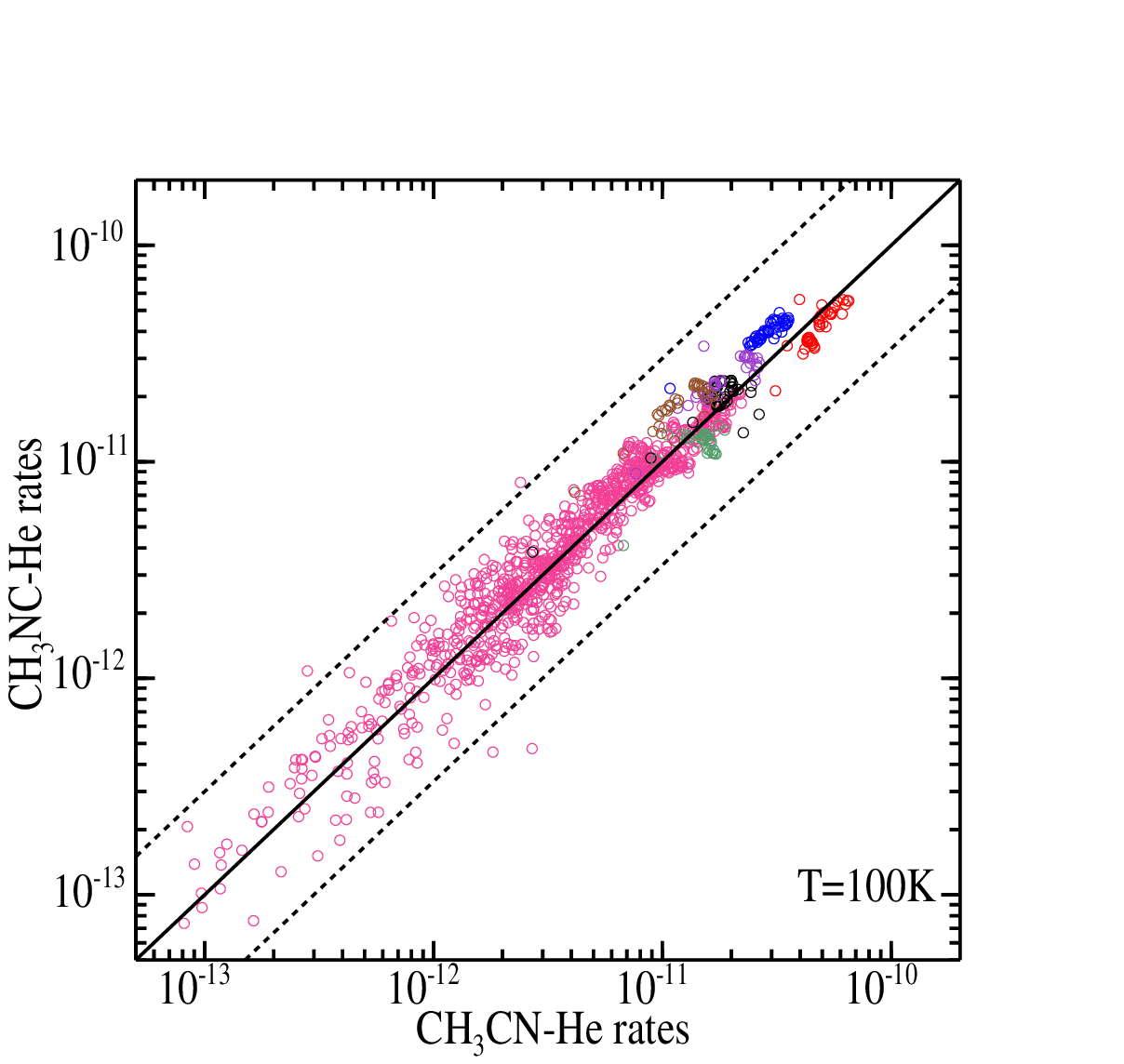}}
{\label{c}\includegraphics[width=.49\linewidth]{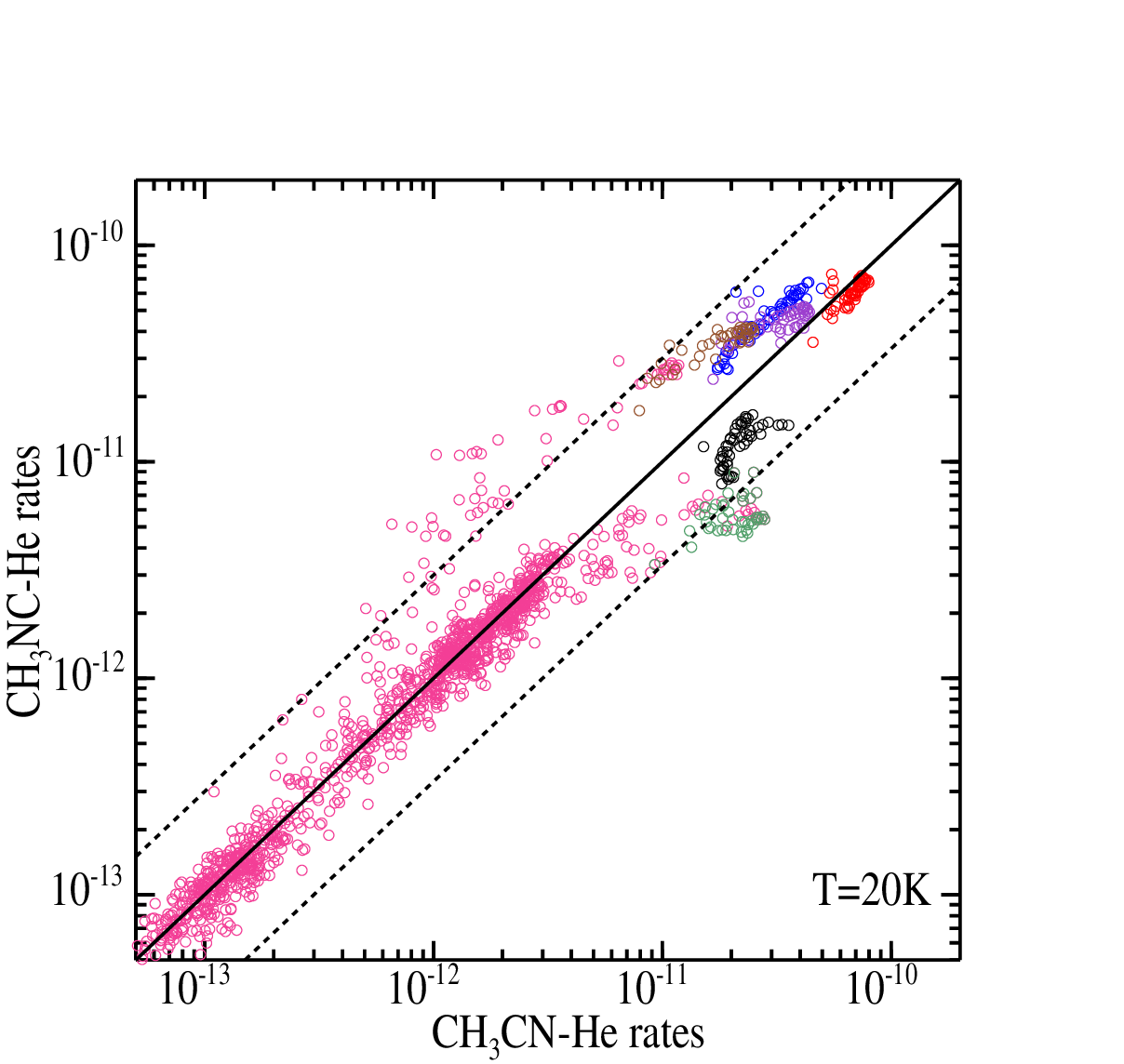}}
{\label{c}\includegraphics[width=.49\linewidth]{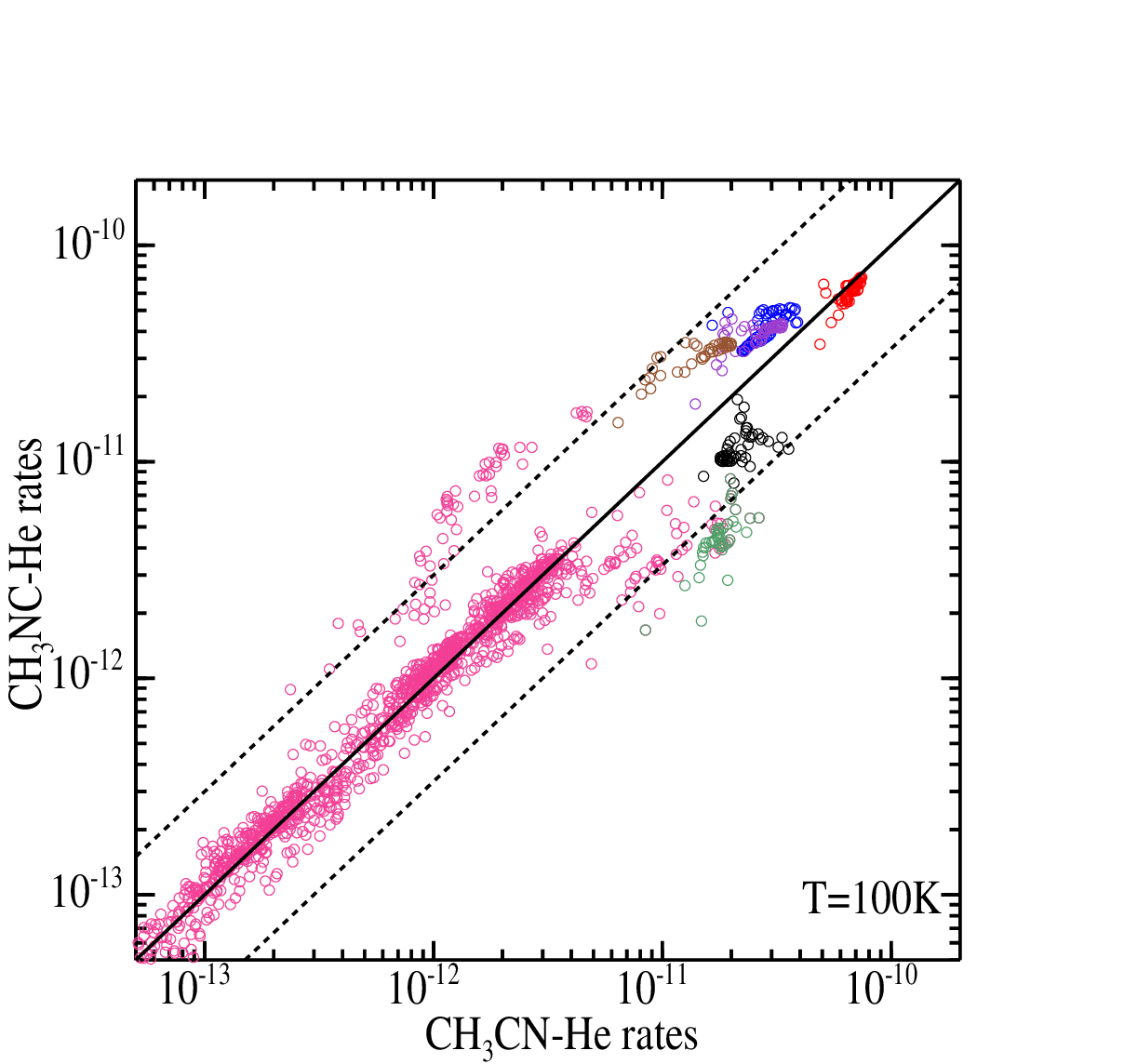}}
\caption{Comparison between $ortho$ (top panels) and $para$ (bottom panels) CH$_3$CN-He and CH$_3$NC-He rate coefficients at $T=20$ K and $T=100$ K. The diagonal line corresponds to equal rate coefficients. Blue circles: transitions $\Delta j=1$, $\Delta k=0$; Red circles: $\Delta j$=2, $\Delta k$=0; Black circles: $\Delta j$=3, $\Delta k$=0; Purple circles: $\Delta j$=4, $\Delta k$=0; Green circles: $\Delta j$=5, $\Delta k$=0; Brown circles: $\Delta j$=6, $\Delta k$=0; Pink circles : all remaining quenching transitions ($\Delta j$=1,2,3,..., $\Delta k$=1,2,3,...)}\label{comparison}
\label{comp} 
\end{figure*}

Because of the absence of rate coefficients for the excitation of CH$_3$NC, the collisional data for CH$_3$CN are often used in radiative transfer models to estimate the abundance of CH$_3$NC in the interstellar medium. In order to examine the validity of such an approach and the impact of isomerism effects in the collisional studies, we present in Fig~\ref{comp} a systematic comparison of CH$_3$CN-He and CH$_3$NC-He rate coefficients, computed using the same level of theory (from \textit{ab initio} to dynamics calculations). The comparison includes collisional rate coefficients for transitions among the first 50 $ortho$ levels and 63 $para$ levels of CH$_3$CN and CH$_3$NC at two temperatures, $T=20$~K and $T=100$~K.

The rate coefficients for the dominant transitions ($k$ > 10$^{-11}$cm$^3$) of the two isomers are within a factor of 2, but significant differences exist for the smaller rate coefficients with relative deviations that can reach a factor of more than 3. 
Given the propensity rules discussed above, it is interesting to compare the excitation of CH$_3$CN and CH$_3$NC in more detail. Figure~\ref{comp} reveals that the dominant transitions all correspond to $\Delta k =0$ transitions. This can be explained by the fact that for symmetric tops these transitions are driven by the terms with $m=0$ in the expansion (\ref{eq_expansion}), which are the largest ones, while transitions with $\Delta k =1$ require terms with $m=3$, which are small \citep{Loreau2015a,ben2022interaction}.
The plots in Fig.~\ref{comp} also show the clear propensity for $\Delta j=2$ transitions in the case of CH$_3$CN-He, while for CH$_3$NC-He the rate coefficients for $\Delta j=1$ and $\Delta j=2$ are of similar magnitude. This reflects the fact that for CH$_3$NC-He, for low $j$ the  $\Delta j=1$ transitions dominate, while the reverse is true for high values of $j$.
Furthermore, the rates coefficients for $\Delta j=2$, $\Delta k=0$ transitions are larger for CH$_3$CN-He than for CH$_3$NC-He while the reverse is true for $\Delta j=1$, $\Delta k=0$ as well as $\Delta j=4$, $\Delta k=0$ and $\Delta j=6$, $\Delta k=0$. The transitions corresponding to $\Delta j=3$, $\Delta k=0$ and $\Delta j=5$, $\Delta k=0$ are similar in the case of $ortho-$CH$_3$CN/NC-He but larger for $para-$CH$_3$CN-He than $para-$CH$_3$NC-He.

The present results confirms that collisional rates should be obtained independently for different isomers and confirms previous results on isomerism effects found for other cyanides and isocyanides, such as HCN/HNC \citep{dumouchel2010rotational}, HC$_3$N/HC$_2$NC/HNC$_3$ \citep{bop2019isomerism} AlCN/AlNC and MgCN/MgNC \citep{hernandez2013cyanide}.
The rate coefficients obtained for the CH$_3$CN and CH$_3$NC isomers allow us to predict that the excitation of these species in the interstellar medium is different and that the abundance ratio of these molecules cannot be obtained by examining only the brightness temperature ratio of the lines. Therefore, a radiative transfer study must be performed for both isomers using their respective set of collisional rates coefficients to obtain the relative abundance of the two isomers with a good precision.

\subsection{Comparison with previous results}
Due to the importance of the CH$_3$CN molecule, and its omnipresence, a set of rate coefficients for the
collisional de-excitation by H$_2$ molecules has been calculated by \citet{green1986collisional}, although using a very approximate treatment of both the \textit{ab initio} potential and of the nuclear dynamics. To the best of our knowledge, these are the only rate coefficients known for this system and used in non-LTE radiative transfer modelling. It is therefore instructive to compare our results to those of  \citet{green1986collisional}. 

\begin{figure*}
\centering
{\label{a}\includegraphics[width=.49\linewidth]{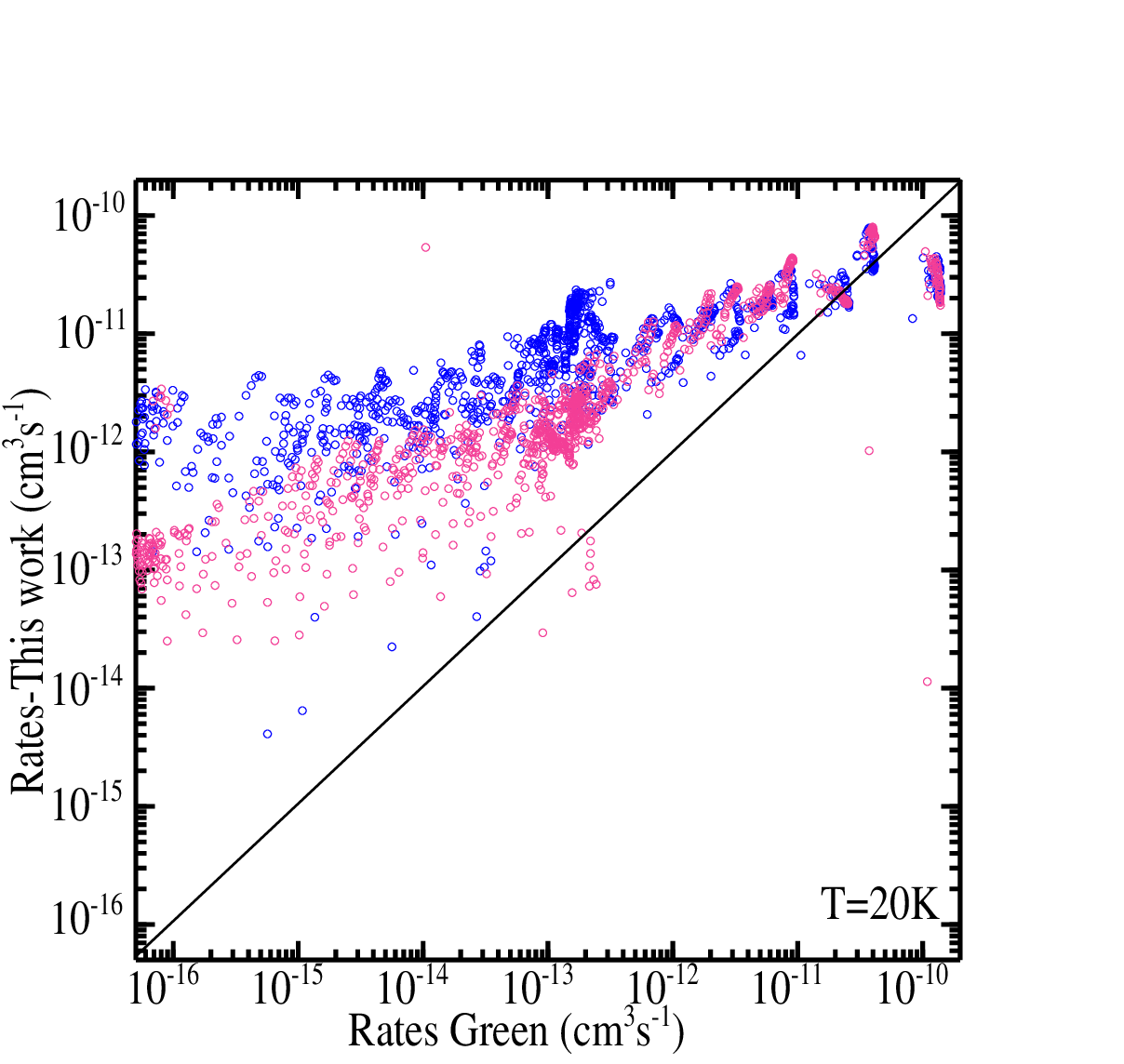}}
{\label{b}\includegraphics[width=.49\linewidth]{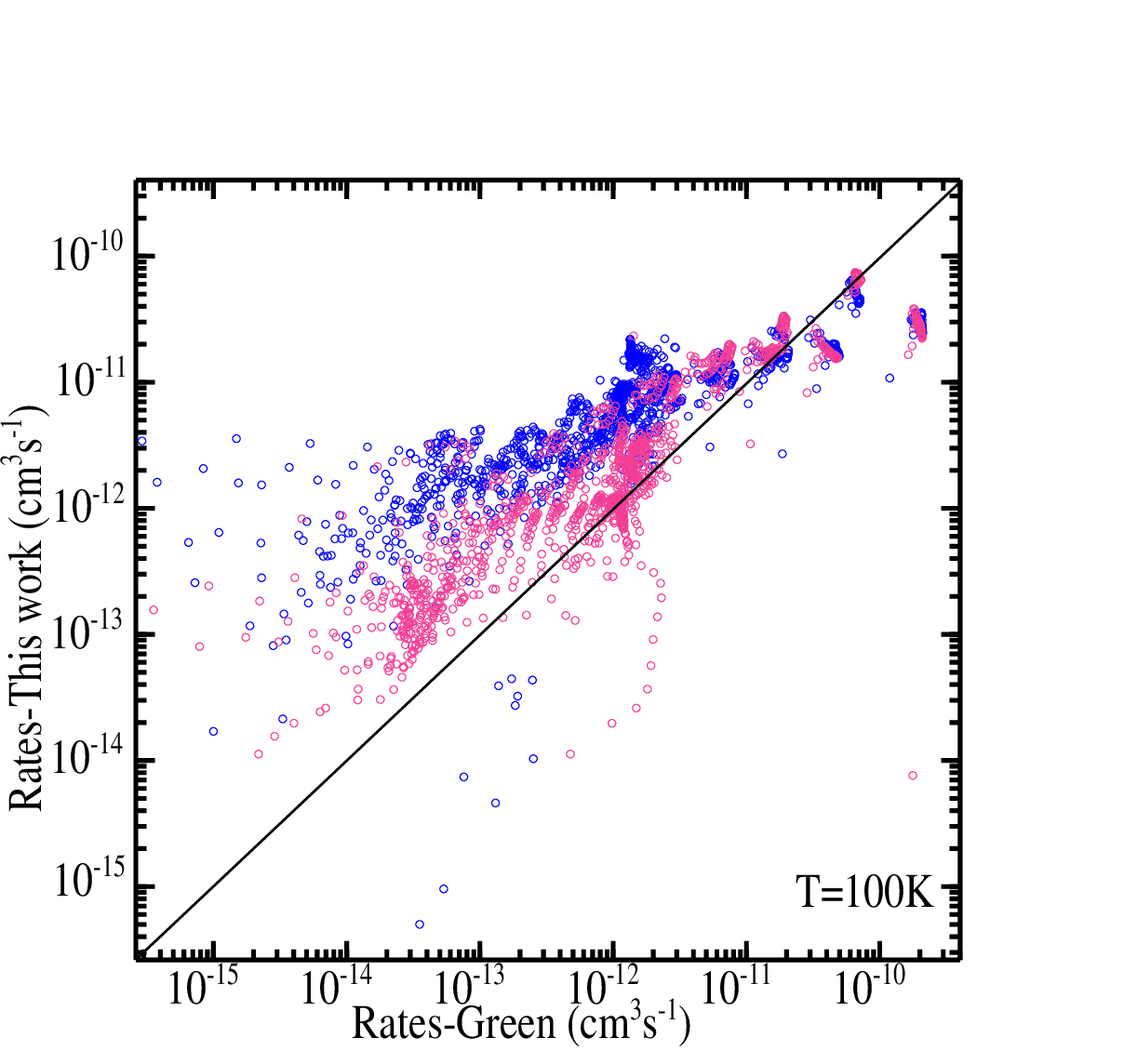}}
\caption{Comparison of quenching rate coefficients obtained in the present work and the published LAMDA rates from \citet{green1986collisional} for transitions among the lowest 50 levels of $ortho-$CH$_3$CN-He (blue circles) and $para-$CH$_3$CN-He (pink circles) at $T=20$ K and $T=100$ K.}
\label{comp-Green} 
\end{figure*}

Because of the spherical character of $para-$H$_2(j=0)$, we compare collisional rates of CH$_3$CN-H$_2$ with those with helium atoms, duly scaled by a factor $\sim$ 1.39 that reflects the difference in reduced masses.
Figure~\ref{comp-Green} highlights the quenching rates computed at temperatures of 20 K and 100 K among the lowest 50 levels of $ortho$-CH$_3$CN-He and the lowest 50 levels of $para$-CH$_3$CN-He. 
We note that the collisional rate coefficients differ greatly, with differences that can reach three orders of magnitude for numerous transitions. The discrepancy is seen to be larger at low temperature (20 K) than at high temperature (100 K). 

Globally, rate coefficients computed by Green are lower than the present results. Two main reasons can explain such large differences: ($i$) the use of two different interaction potentials, the potential of Green having being obtained with an approximate electron gas method; and ($ii$) the treatment of the dynamics. The present work uses either an accurate quantum approach (CC) or an approximate quantum approach (CS) whose validity has been tested for the present system. The rate coefficients computed by Green were obtained within the infinite order sudden (IOS) approximation for only 12 collision energies, which does not appear to be accurate in the case of CH$_3$CN-He. The difference in accuracy between the CS and IOS methods is expected to decrease with increasing energy, which likely explains the slightly better agreement at 100 K compared to 20 K.
The large difference between collisional rates computed by Green compared to our CH$_3$CN-He rates calls for a re-examination of their impact on the abundance of CH$_3$CN in various astrophysical environments.

\section{Astrophysical application}\label{astro}

The comparison of collisional rate coefficients performed in section~\ref{Tauxx} showed that CH$_3$CN and its isomer CH$_3$NC have different excitation behaviors when they collide with helium atoms. It is therefore instructive to study the impact of using the appropriate rate coefficients when deriving the abundance of CH$_3$CN and CH$_3$NC in astrophysical environments. 

As a first application, we performed a radiative transfer computation to simulate the excitation of both isomers within the escape probability formalism approximation for a uniform spherical geometry of a homogeneous interstellar medium as implemented in the RADEX code \citep{van2007computer}. The molecular data (for CH$_3$CN and CH$_3$NC) are composed of rate coefficients (from 5 to 100K) scaled from He to H$_2$ and supplemented by the energy levels, Einstein coefficients, as well as line frequencies of both isomers. These spectroscopic data were obtained from the Cologne Database for Molecular Spectroscopy (CDMS) portal \citep{endres2016cologne}.\\
We first computed the critical density of colliders $n^*(H_2,T)$ using the equation:
\begin{equation}
n^*(H_2,T)=\frac{A_{fi}}{\sum_{f \neq i}k_{fi}}
\end{equation}
where $k_{fi}$ and $A_{fi}$ are collisional rate coefficients and Einstein spontaneous absorption coefficients, respectively.

The critical density was computed for transitions involving the emission in the range of 91.98 GHz and 220.74 GHz for CH$_3$CN and 100.52 GHz and 241.23 GHz for CH$_3$NC, covering lines that are often detected with ground based radio telescopes. This range of frequency includes the often detected transitions $j_k$= 5$\rightarrow$4 and 12$\rightarrow$11 with $k$ or $k^\prime= 0,1,2$ and 3. The critical densities of colliders (in cm$^{-3}$), computed at $T =$10, 50 and 80K, are given in table~\ref{density}.
It is noticeable that the critical density of these observed transitions is around $\sim$10$^4$ cm$^{-3}$, which corresponds to the typical density of the interstellar medium. Therefore, the local thermodynamic equilibrium cannot be assumed to be reached. Consequently, the use of non-LTE simulations is important in order to accurately model the CH$_3$CN and CH$_3$NC emission spectra.

We calculated the brightness temperature ($T_B$) and the
excitation temperature (T$_{\textrm{ex}}$) for various pure rotational emission lines which are frequently observed in astrophysical clouds. We take here as examples the $j_k$=2 $\rightarrow$ 1, 5 $\rightarrow$ 4 and 6 $\rightarrow$ 5 with $k$= 0,1,2 and 3 transitions.

In the radiative transfer computation, the cosmic microwave background  was introduced as a background radiation field ($T_{\textrm{CMB}}$=2.73 K). The line width at the half-maximum (FWHM) 
was fixed at 25.0, 6.0 and 9.2 km.s$^{-1}$ for $j_k$=2 $\rightarrow$ 1, 5 $\rightarrow$ 4 and 6 $\rightarrow$ 5 respectively \citep{blackman1977detection,araya2005ch3cn}, and the kinetic temperature was set at 20, 50 and 80 K.

In order to cover the H$_{2}$ volume density [$n$(H$_{2}$)] and to check the behavior of CH$_3$CN and CH$_3$NC excitation under and out of LTE conditions, we varied $n$(H$_{2}$) from  10$^{2}$ cm$^{-3}$ to 10$^8$ cm$^{-3}$ and the column densities of CH$_3$CN and CH$_3$NC were fixed  at $N_c$=10$^{14}$ cm$^{-2}$, a choice that is based on the molecule's estimated column density of both isomers in the ISM \citep{araya2005ch3cn}. However, a variation up to an order of magnitude of this quantity does not affect the magnitude of the excitation temperatures because of the optically thin regime.

\begin{figure}
\centering
\includegraphics[width=0.9\columnwidth]{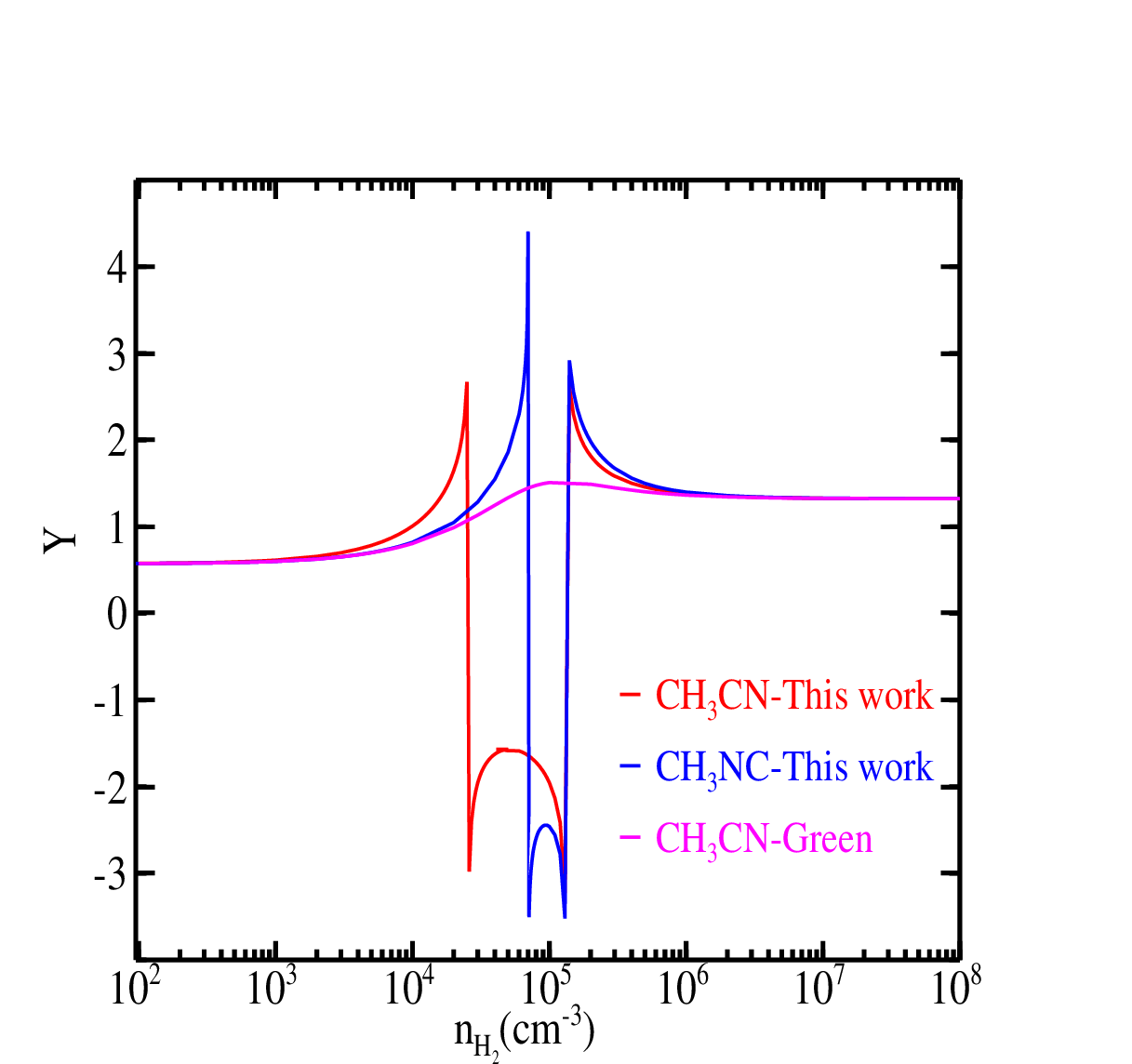}
	\caption{Variation of the logarithm of the excitation temperature as a function of the H$_2$ density for the $j_k$= 2$_1 \rightarrow 1_1$ CH$_3$CN and CH$_3$NC rotational transitions at kinetic temperature of 20 K. In order to better describe the large amplitude of its variation, the excitation temperature is represented on this figure as
 $Y=\frac{T_{ex}}{|T_{ex}|} \times log_{10}(1+|T_{ex}|)$. The results from the present work are compared to those obtained with the rate coefficients of \citet{green1986collisional}. }
	\label{neg}
\end{figure}

Figure~\ref{neg} presents the logarithm of the excitation temperature obtained for the transition $j_k$= 2$_1 \rightarrow 1_1$ using rate coefficients of $para$ CH$_3$CN/NC computed during this work and CH$_3$CN rates computed by Green at $T=20$ K. As can be seen in the figure, a population inversion is present for both isomers, which depends on the H$_2$ density, when we use the new set of collisional rates. One can see that the excitation temperature is negative in the H$_2$ density range of 10$^4$-10$^6$ cm$^{-3}$. This negative excitation temperature behavior indicates a population inversion ($n_u$/$g_u$ > $n_l$/$g_l$). 
However, this behavior is not predicted using the rate coefficients from the literature. The excitation temperature increases smoothly and tends towards the kinetic temperature, and no population inversion is observed when we use rates of Green. Consequently, a reexamination of the temperatures assigned to low density gases might be required.  

\begin{figure*}
\centering
{\label{a}\includegraphics[width=.33\linewidth]{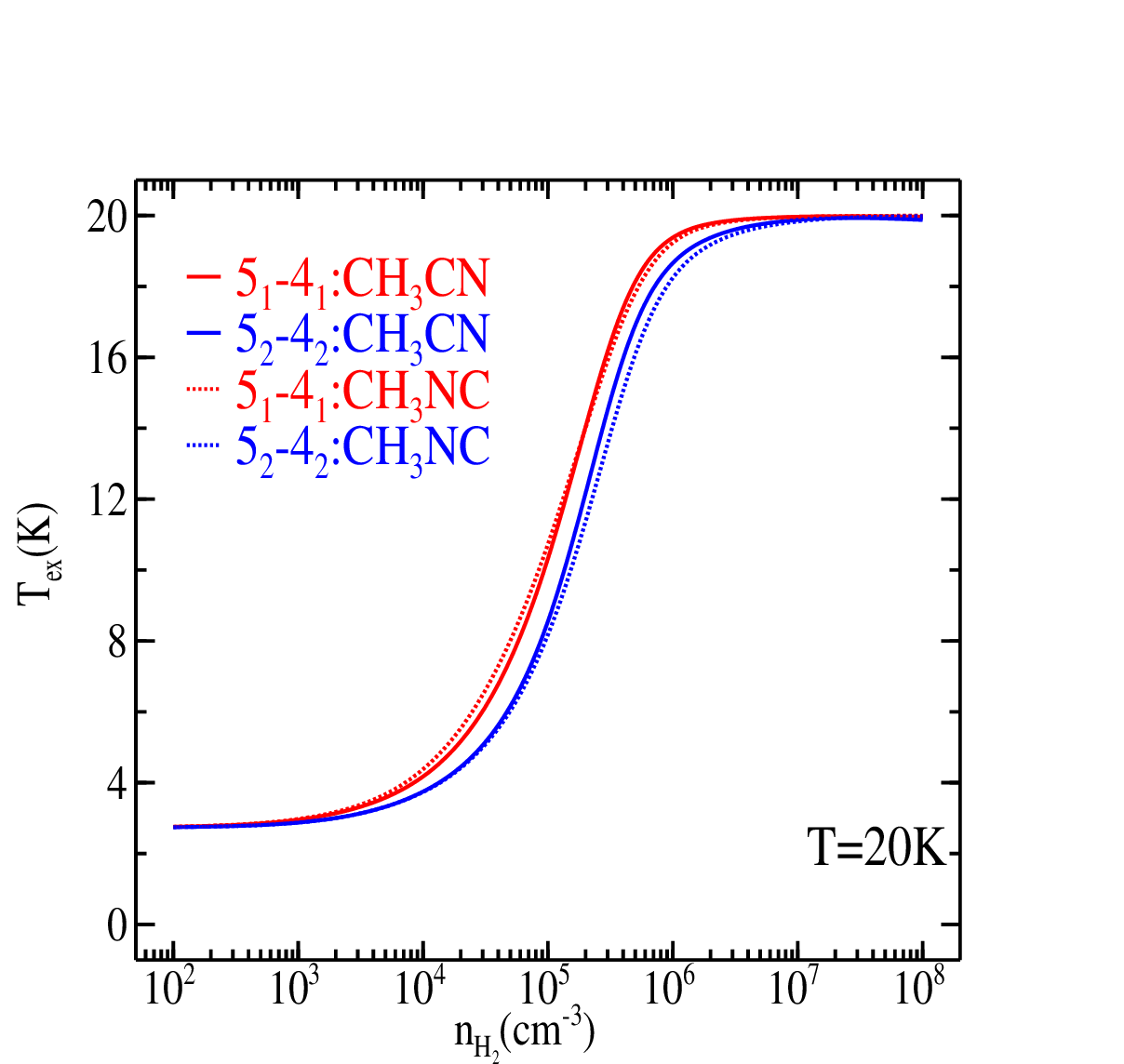}}
{\label{b}\includegraphics[width=.33\linewidth]{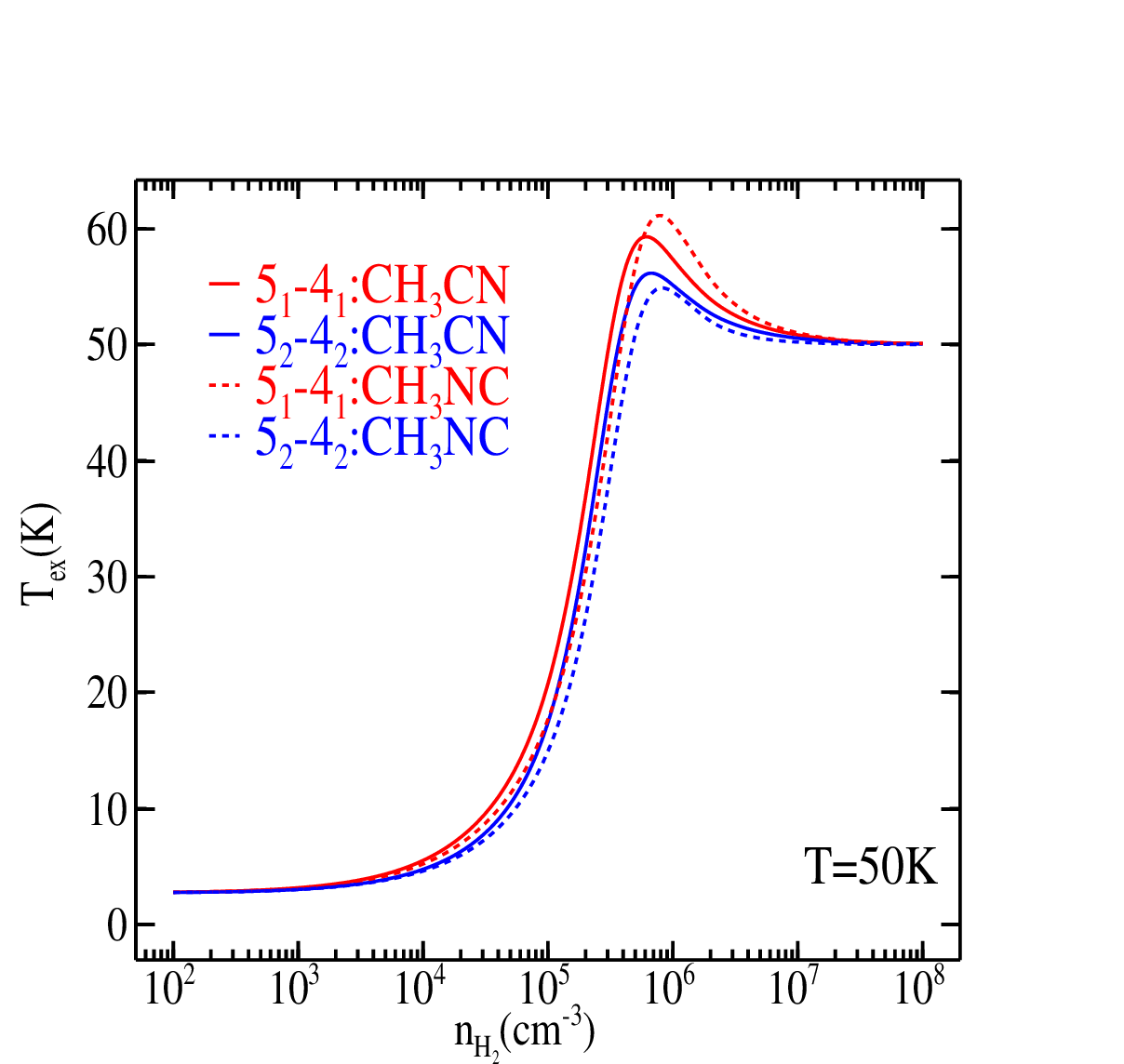}}
{\label{c}\includegraphics[width=.33\linewidth]{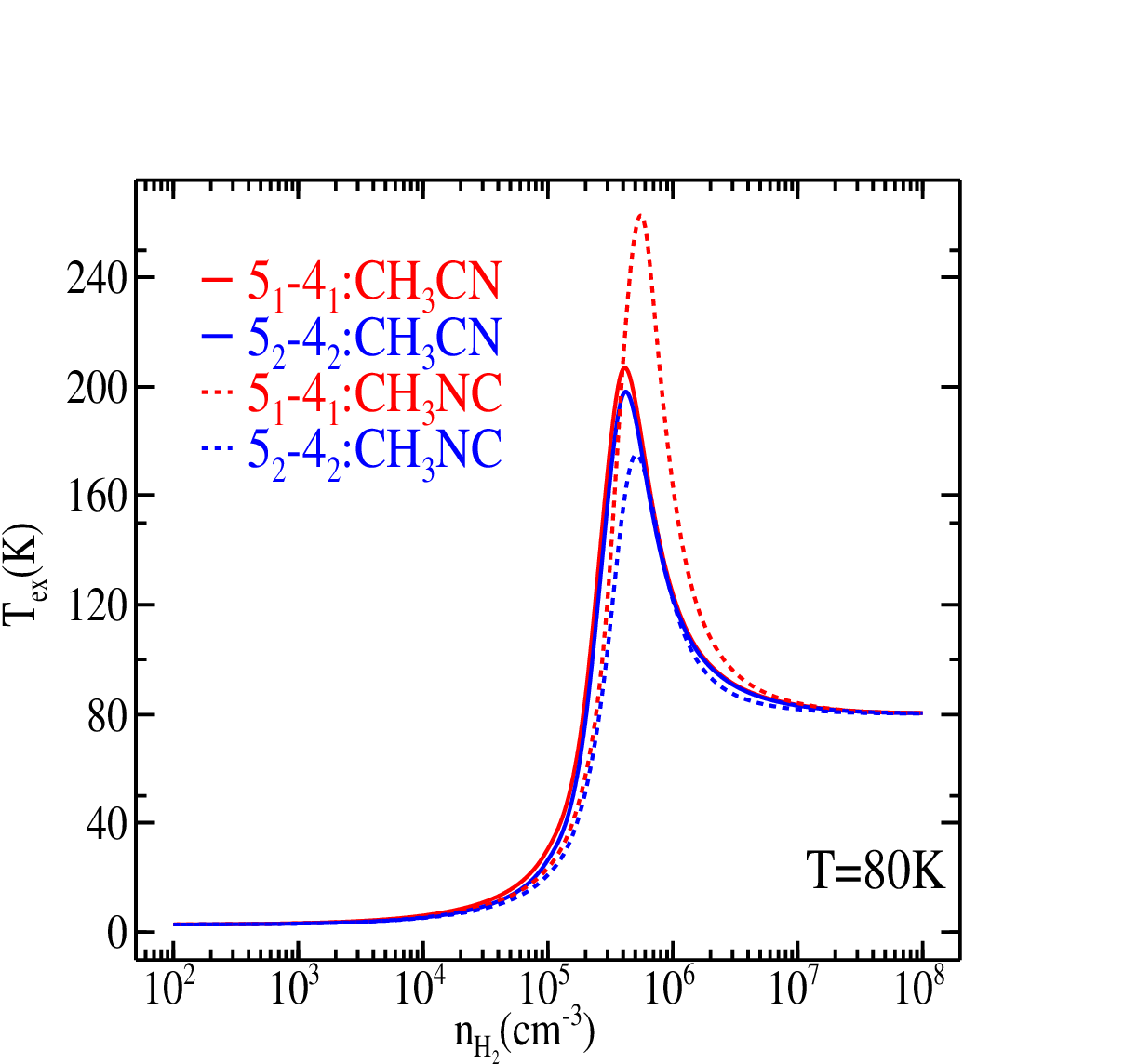}}
{\label{c}\includegraphics[width=.33\linewidth]{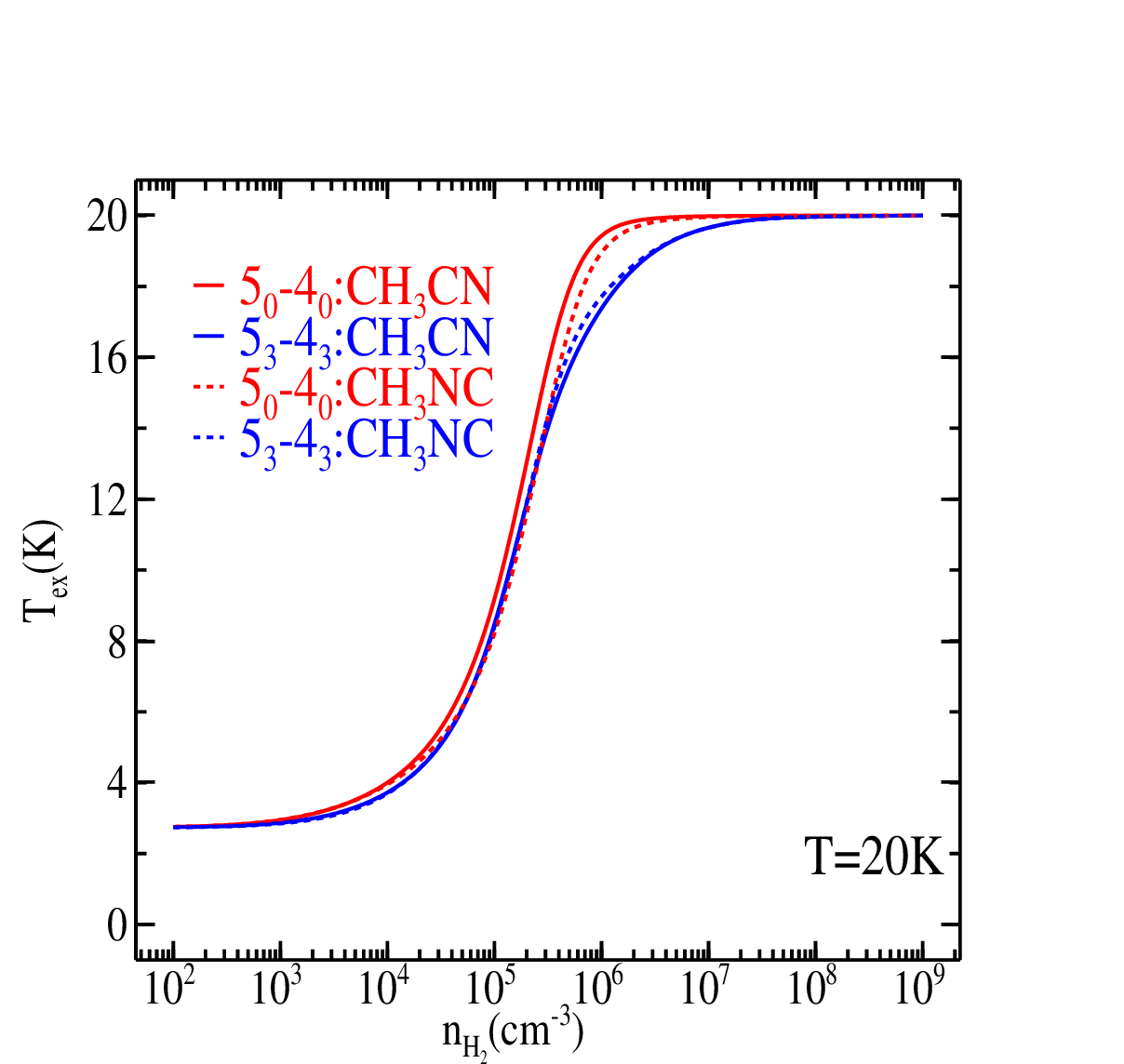}}
{\label{c}\includegraphics[width=.33\linewidth]{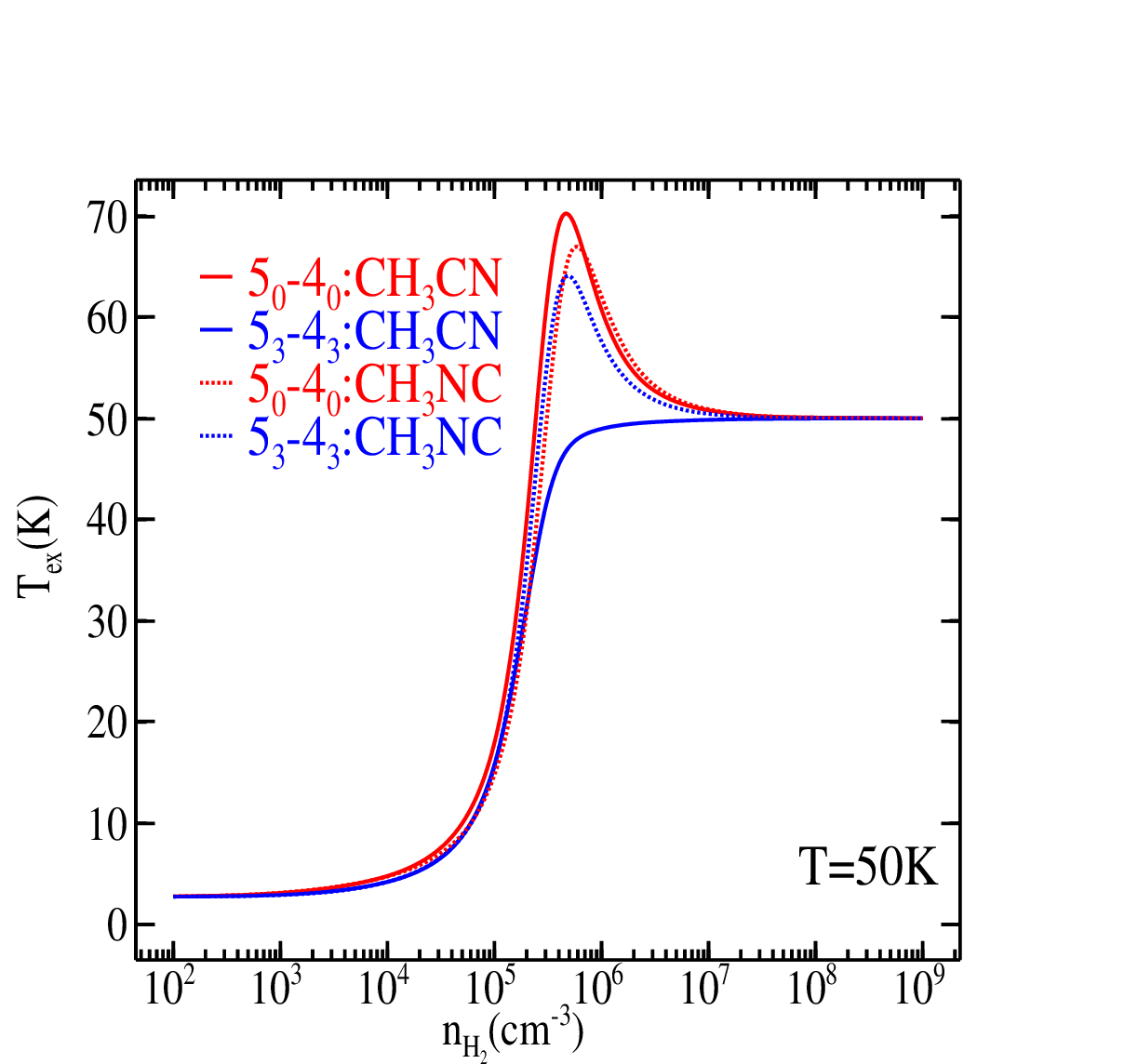}}
{\label{c}\includegraphics[width=.33\linewidth]{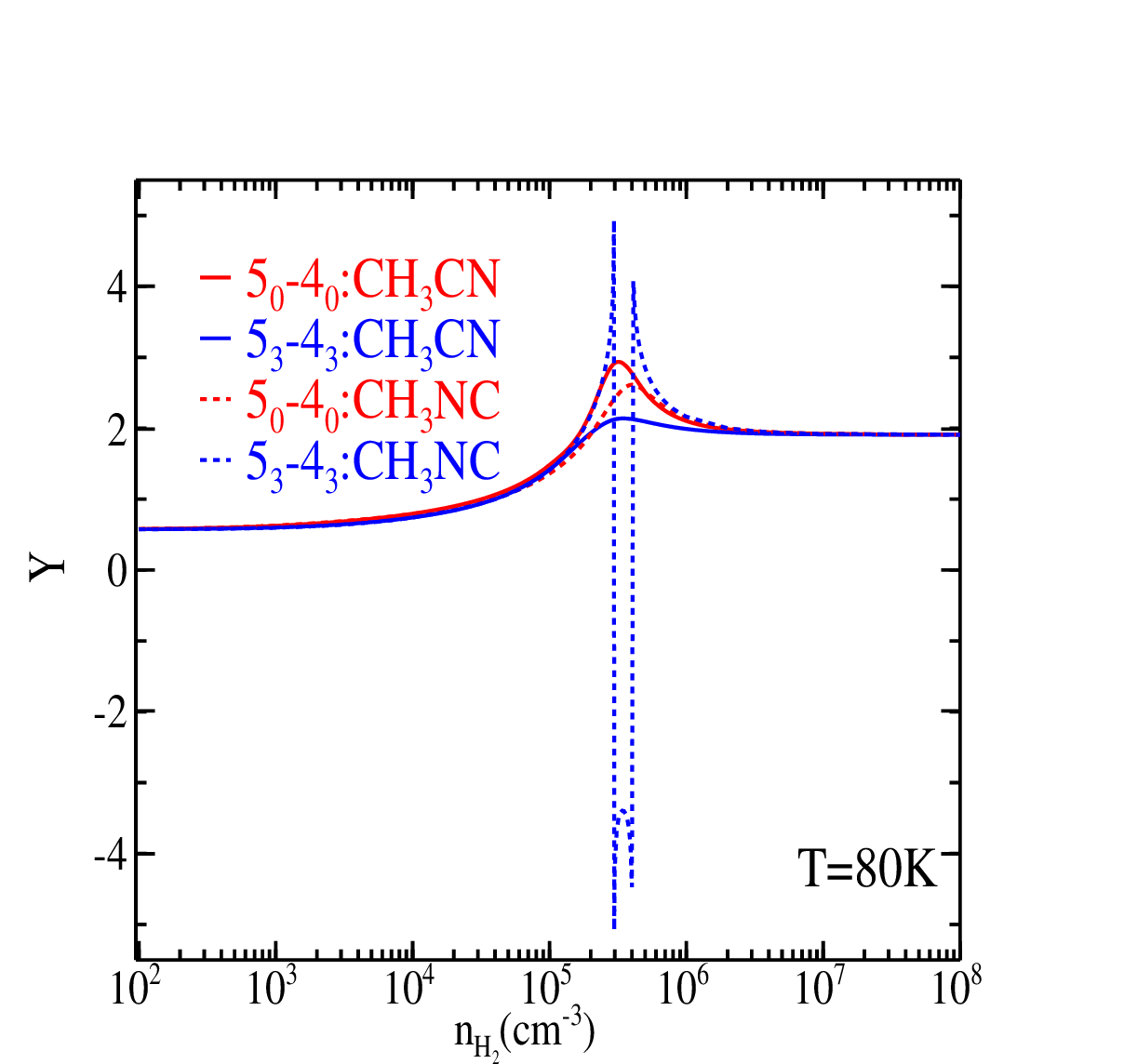}}
\caption{Excitation temperature of CH$_3$CN (solid curves) and CH$_3$NC (dashed curves) for the transitions
$j_k \rightarrow j^\prime_{k^\prime}$ with $j=5$, $j^\prime=4$ and $k=k^\prime=1$ or 2 ($para$-CH$_3$CN/NC, top panels) and $k=k^\prime=0$ or 3 ($ortho$-CH$_3$CN/NC, bottom panels)
as a function of the H$_2$ density for three kinetic temperature (20 K, 50 K and 80 K) and a column density of 10$^{14}$ cm$^{-2}$. The bottom right panel shows $Y=\frac{T_{ex}}{|T_{ex}|} \times log_{10}(1+|T_{ex}|)$ instead of the excitation temperature.}
\label{application}
\label{application-ex} 
\end{figure*}

Figure~\ref{application-ex} displays the excitation temperature obtained for the transitions $j_k$= 5 $\rightarrow$ 4 with $k$= 0,1,2 and 3. At low H$_{2}$ density, the medium is dilute and the excitation temperature is equal to $T_{\textrm{CMB}}$=2.73K, a value that corresponds to the background radiation field. It increases monotonically as collisional processes become more important predicting no suprathermal excitation at $T=20$K. 
For densities above $n_{\textrm{H$_{\textrm{2}}$}}=10^6$cm$^{-3}$, the excitation temperature tends towards
the kinetic temperature as LTE conditions are reached. At this stage, the populations of the rotational levels no longer depend on the density of H$_2$ and simply obey Boltzmann's law.
Furthermore, at T= 50K and 80K, we note a particular suprathermal excitation for all transitions for H$_2$ densities between 10$^{5}$ and 10$^6$ cm$^{-3}$. This behavior is defined by an excitation temperature (T$_{\textrm{ex}}$) that is greater than the kinetic temperature (T$_k$). For the transition $j_k$= 5$_3 \rightarrow 4_3$ of CH$_3$NC we observe a level population inversion where the excitation temperature becomes negative in approximately the same H$_2$ density range. 

If we analyze the intermediate domain where collisional and radiative processes are in competition, we see that the comparison of the $T_{\textrm{ex}}$ of CH$_3$CN and CH$_3$NC isomers shows that the excitation conditions of both isomers are different and that a similar $T_{\textrm{ex}}$ for both isomers cannot be assumed. 

Figure~\ref{application} presents the brightness temperature of CH$_3$CN and its isomer CH$_3$NC as a function of the H$_2$ volume density.
In all panels, the brightness temperature globally increases with the increase of the density and the temperature with an asymptotic behavior at $n$ > 10$^6$ cm$^{-3}$. A small peak is observed for $T =50$ K and $5\times10^{4} \leq n_{H_2} \leq 5\times 10^{5}$ cm$^{-3}$ with a higher magnitude at $T=80$ K.
We can see several differences between the two isomers, especially in the intermediary domain ($10^4 \leq  n \leq 10^6$ cm$^{-3}$) where collisional and radiative processes are in competition. These differences demonstrate the need of calculating collision rate coefficients of each isomer separately to determine their abundance with good precision.

\begin{table}
\centering
\caption{CH$_3$CN and CH$_3$NC critical densities $n_c^*$(cm$^{-3}$) at 10, 50 and 80 K for several observed transitions.}\label{density}
\begin{tabular}{ccccc}
\hline
\hline
Molecule &  Transition & $T=10$ K & $T=50$ K & $T=80$ K \\
\hline
CH$_3$CN & 5$_0$ $\rightarrow$ 4$_0$ & $0.3768 \times 10^4$ & $0.3976 \times 10^4$ &  $0.4021 \times 10^4$ \\
& 5$_1$ $\rightarrow$ 4$_1$ & $0.3732 \times 10^4$ & $0.3940 \times 10^4$ &  $0.4001 \times 10^4$ \\
& 5$_2$ $\rightarrow$ 4$_2$ & $0.3731 \times 10^4$ & $0.3939 \times 10^4$ &  $0.4001 \times 10^4$ \\
\hline
& 12$_0$ $\rightarrow$ 11$_0$ & $0.5568 \times 10^5$ & $0.5889 \times 10^5$ &  $0.6097 \times 10^5$ \\
& 12$_1$ $\rightarrow$ 11$_1$ & $0.6284 \times 10^5$ & $0.6487 \times 10^5$ &  $0.6592 \times 10^5$ \\
& 12$_2$ $\rightarrow$ 11$_2$ & $0.6283 \times 10^5$ & $0.6485 \times 10^5$ &  $0.6590 \times 10^5$ \\

\hline
CH$_3$NC & 5$_0$ $\rightarrow$ 4$_0$ & $0.4984 \times 10^4$ & $0.5037 \times 10^4$ &  $0.5118 \times 10^4$ \\
& 5$_1$ $\rightarrow$ 4$_1$ & $0.5492 \times 10^4$ & $0.5547 \times 10^4$ &  $0.5684 \times 10^4$ \\
& 5$_2$ $\rightarrow$ 4$_2$ & $0.4802 \times 10^4$ & $0.4924 \times 10^4$ &  $0.5046 \times 10^4$ \\
 \hline
& 12$_0$ $\rightarrow$ 11$_0$ & $0.7128 \times 10^5$ & $0.7151 \times 10^5$ &  $0.7266 \times 10^5$ \\
& 12$_1$ $\rightarrow$ 11$_1$ & $0.8478 \times 10^5$ & $0.8520 \times 10^5$ &  $0.8610 \times 10^5$ \\
& 12$_2$ $\rightarrow$ 11$_2$ & $0.8333 \times 10^5$ & $0.2487 \times 10^5$ &  $0.8543 \times 10^5$ \\
\hline
\hline
\end{tabular}
\end{table}


\begin{figure*}
\centering
{\label{a}\includegraphics[width=.33\linewidth]{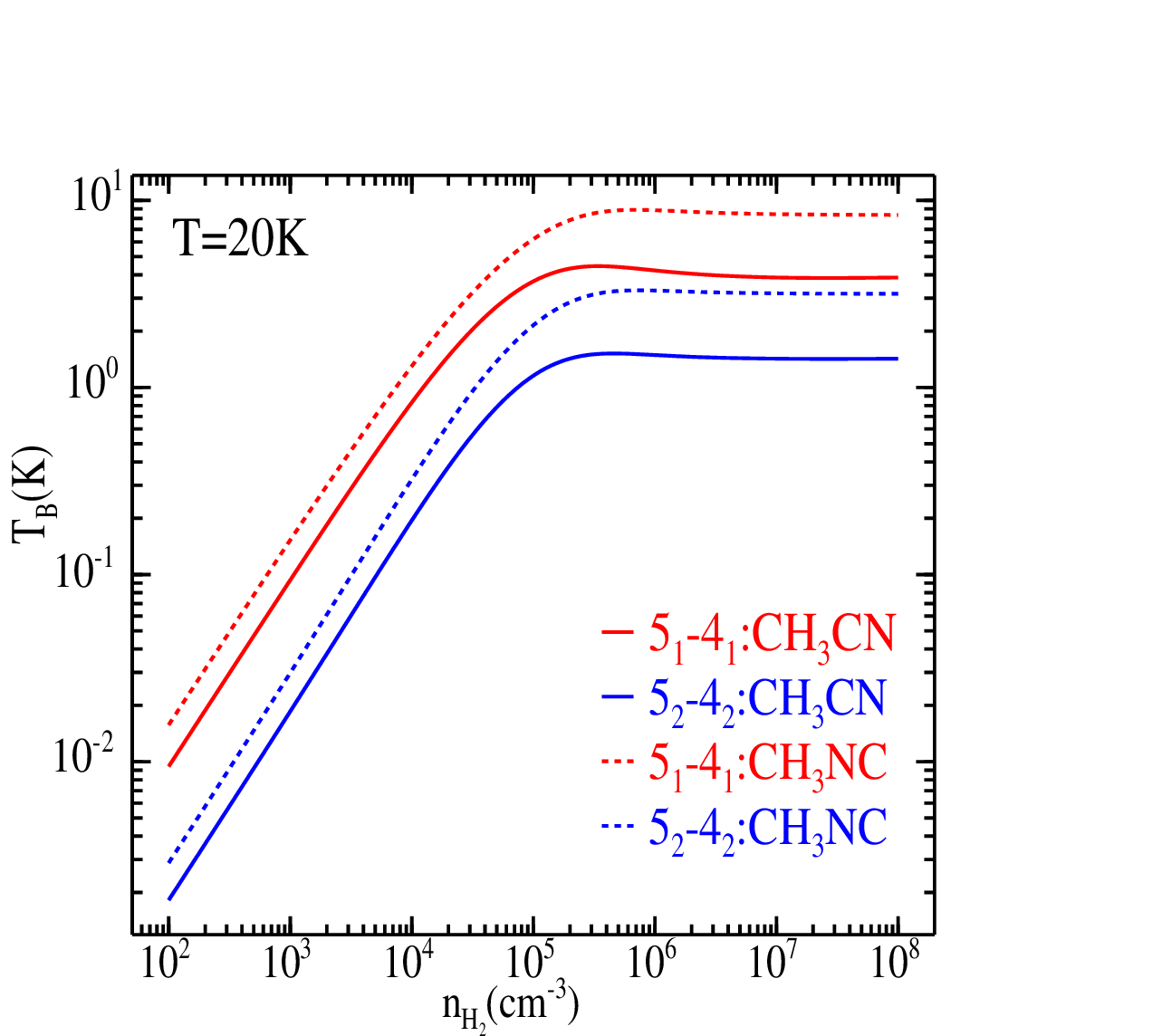}}
{\label{b}\includegraphics[width=.33\linewidth]{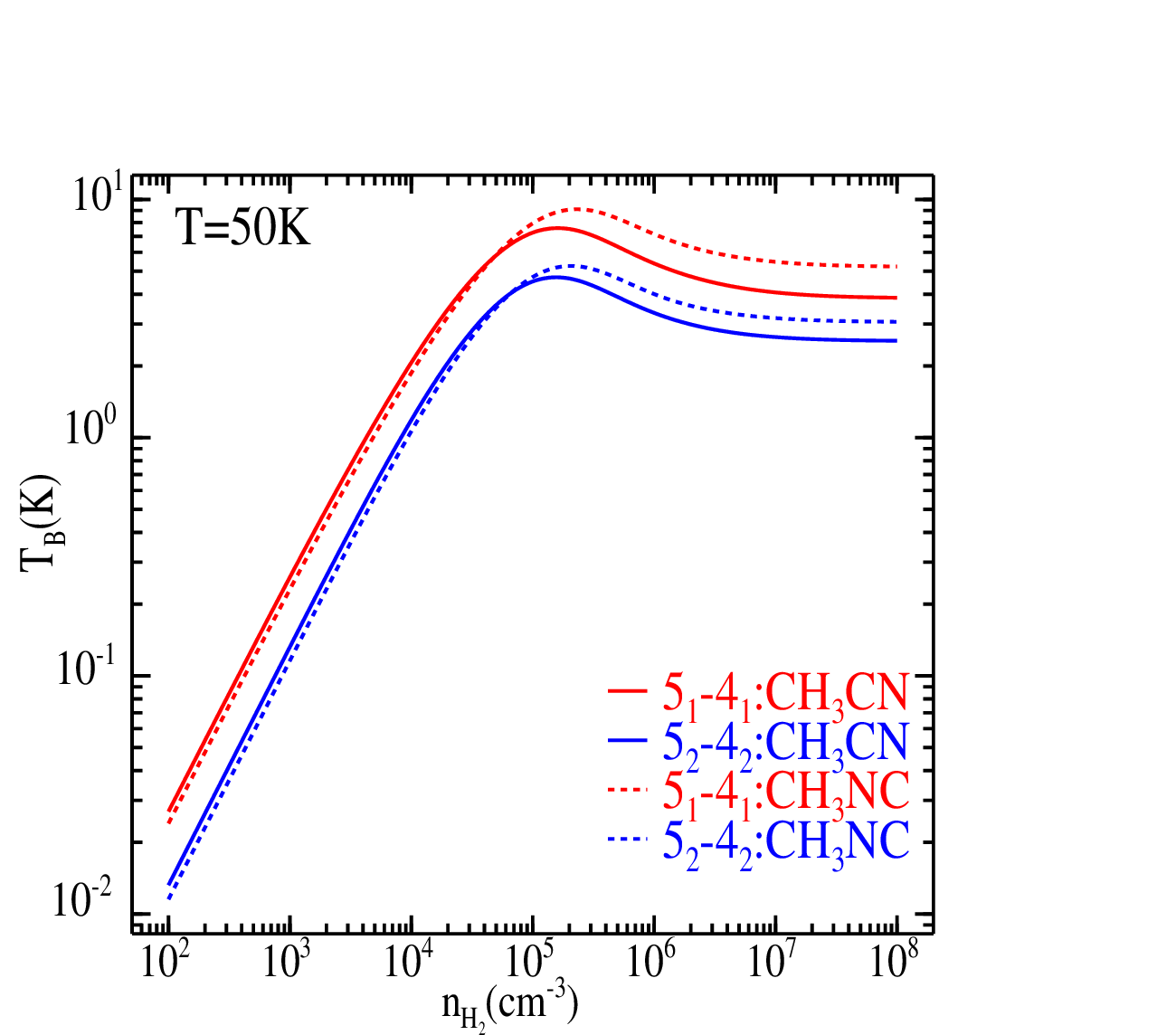}}
{\label{c}\includegraphics[width=.33\linewidth]{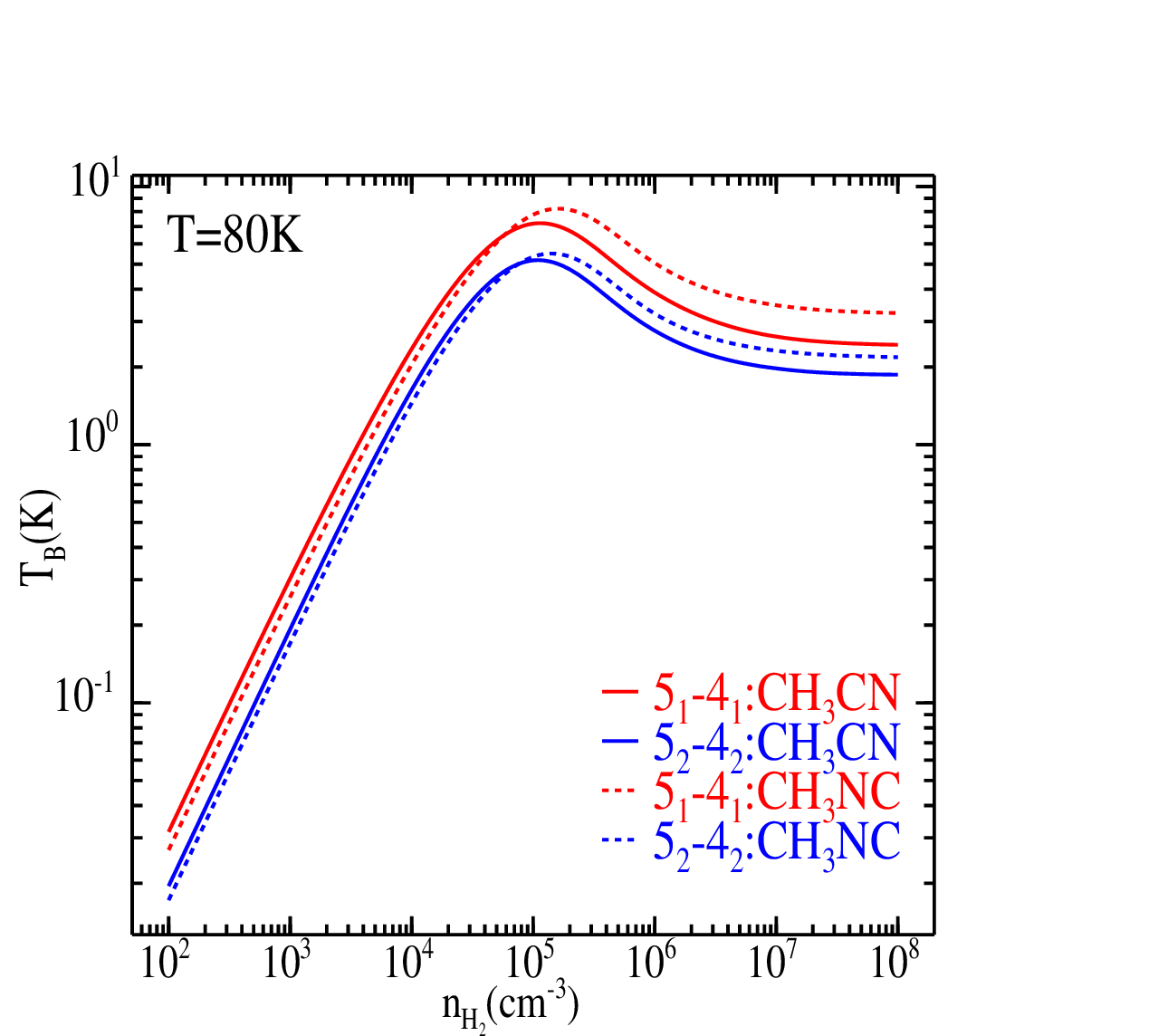}}
{\label{c}\includegraphics[width=.33\linewidth]{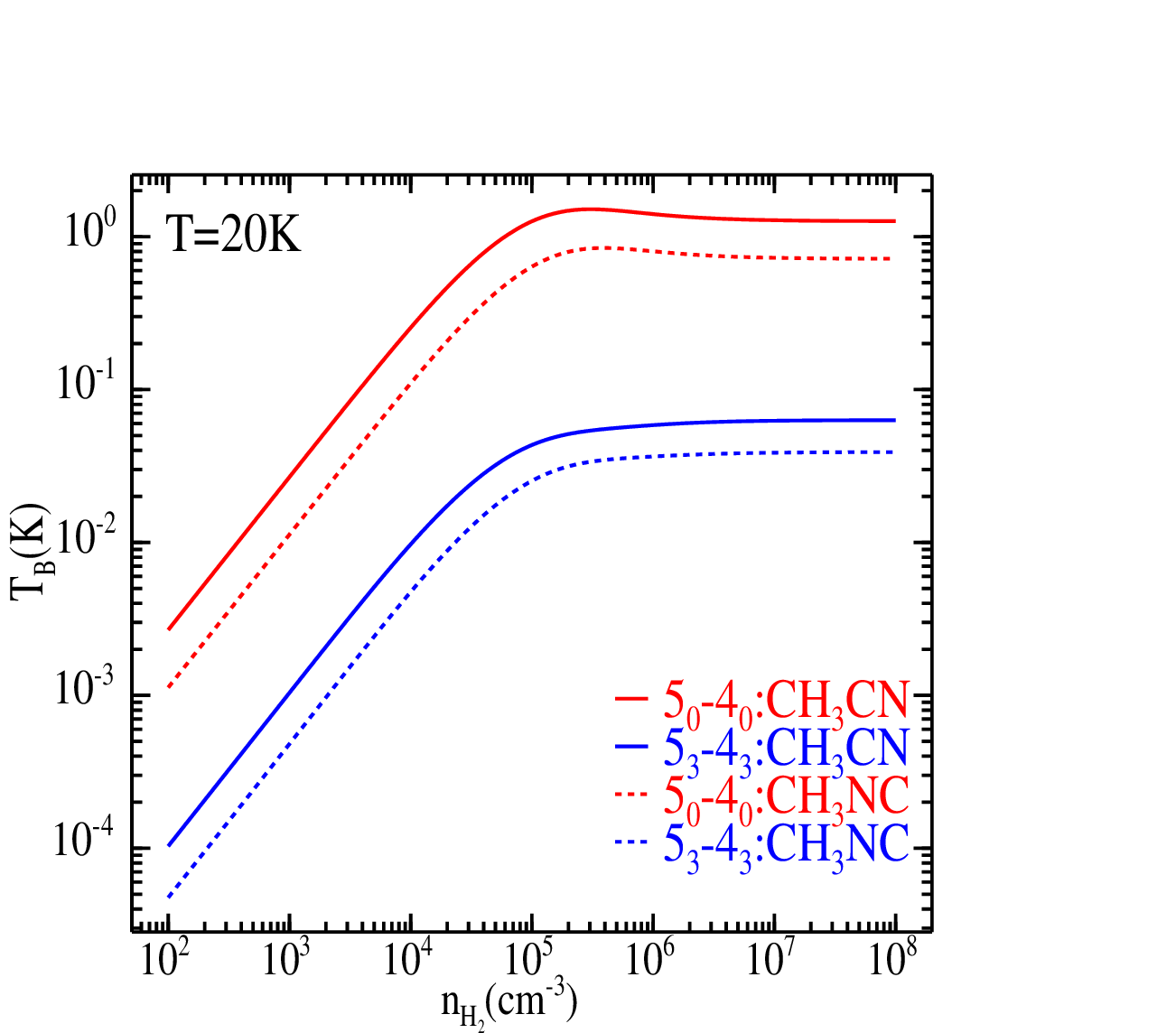}}
{\label{c}\includegraphics[width=.33\linewidth]{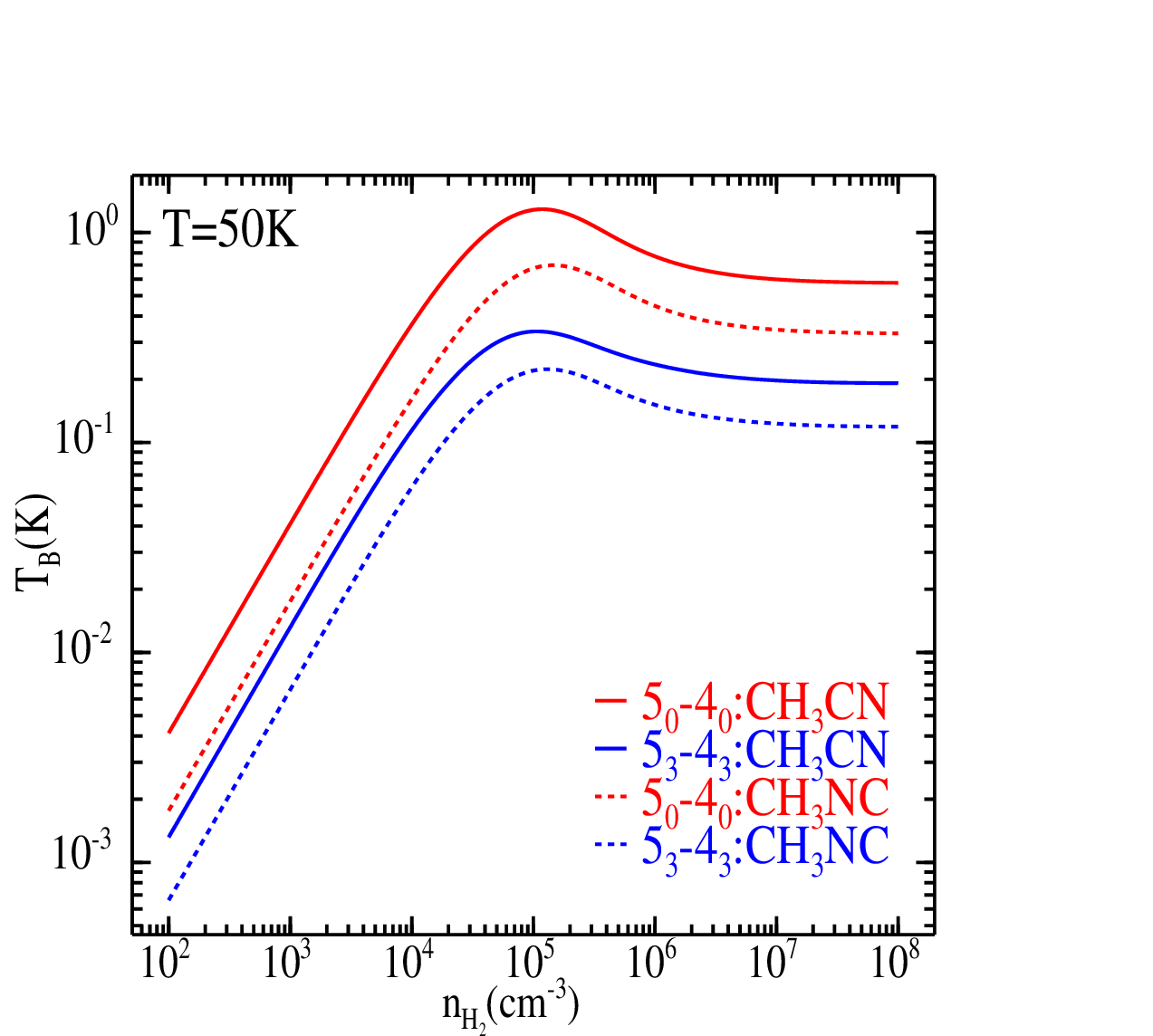}}
{\label{c}\includegraphics[width=.33\linewidth]{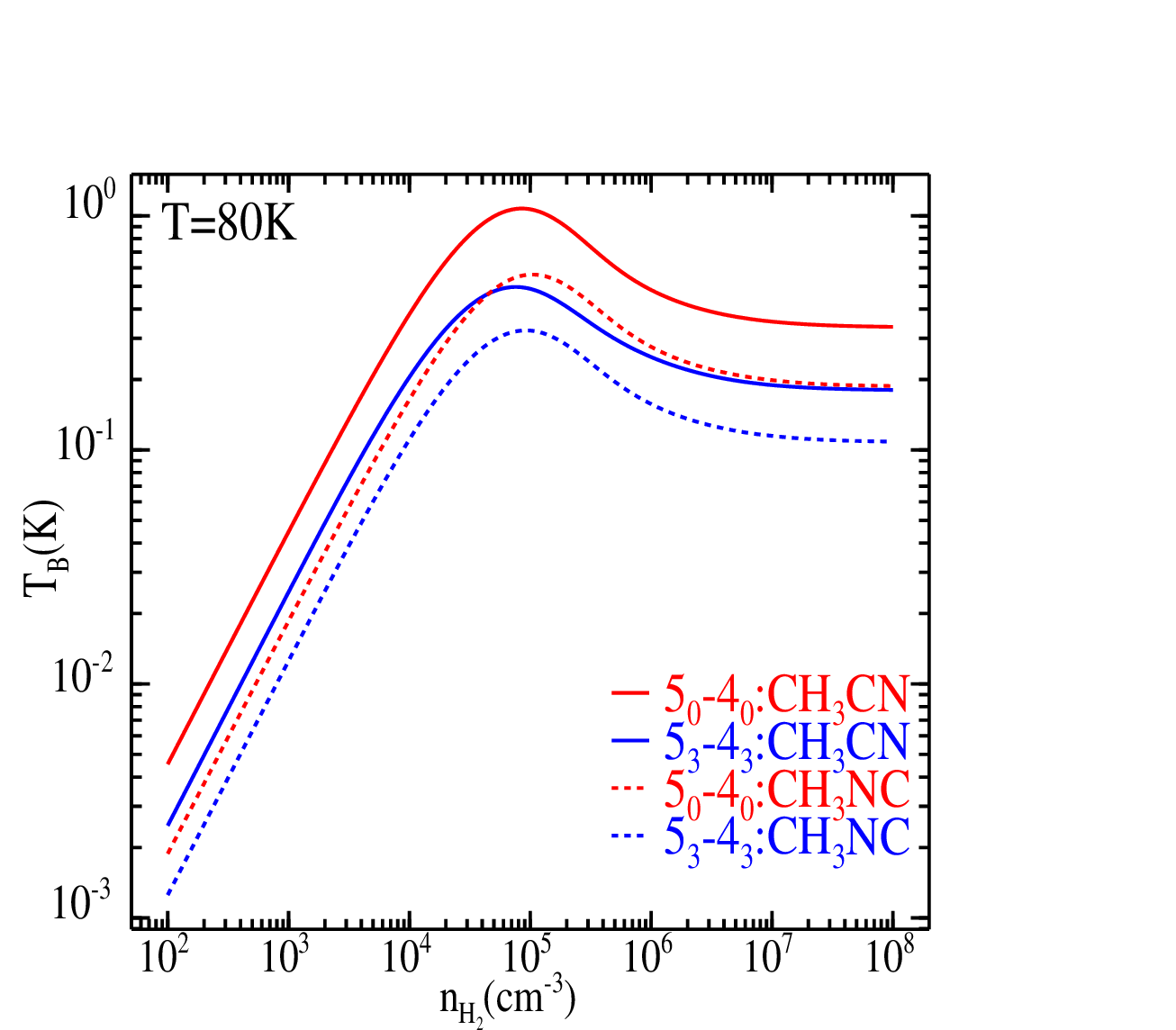}}
\caption{Brightness temperature of CH$_3$CN (solid curves) and CH$_3$NC (dashed curves) for the transitions
$j_k \rightarrow j^\prime_{k^\prime}$ with $j=5$, $j^\prime=4$ and $k=k^\prime=1$ or 2 ($para$-CH$_3$CN/NC, top panels) and $k=k^\prime=0$ or 3 ($ortho$-CH$_3$CN/NC, bottom panels) as a function of the H$_2$ density for three kinetic temperature (20K, 50K and 80K) and a column density of 10$^{14}$ cm$^{-2}$.}\label{application}
\label{application} 
\end{figure*}

\section{conclusion}\label{ccl}
State-to-state cross sections for the (de-)excitation of CH$_3$CN and its isomer CH$_3$NC induced by collisions with helium atoms were computed using the CC method for $E_{\textrm{tot}} \leq$ 100 cm$^{-1}$ and CS approximation for 100 $\leq$ $E_{\textrm{tot}} \leq$ 900 cm$^{-1}$. The dynamics was studied using recent accurate \text{ab initio} PESs. These cross sections were then  averaged over a Maxwell-Boltzmann distribution of velocities  to determine quenching collision rate coefficients between the first 74 levels of $para$-CH$_3$CN/NC and first 66 levels of $ortho$-CH$_3$CN/NC for temperatures ranging from 5 to 100 K. The rate coefficients for the two isomers display different propensity rules. Transitions with $\Delta j=2, \Delta k=0$ are dominant for CH$_3$CN-He collisions, while for CH$_3$NC-He a propensity rule is found in favour of $\Delta j=1, \Delta k=0$ transitions at low temperatures but $\Delta j=2, \Delta k=0$ at high temperatures.  
Overall the CH$_3$CN-He collision rate coefficients are found to be slightly smaller than those for CH$_3$NC-He isomer, which can be explained based on the depth of the potential energy surface corresponding to the collision, which is larger for CH$_3$NC-He than for CH$_3$CN-He (See paper I). 

We have compared the new set of rate coefficients of CH$_3$CN-He with those available in literature for CH$_3$CN-H$_2$ that are currently used for astrophysical modeling. An important discrepancy was found between both sets of data, with differences of up to 3 orders of magnitude for some transitions. There does not seem to be any correlation between the two data sets, which demonstrates the usefulness of the present collisional rates. Even though the present calculations have been performed for the excitation with He atoms and not H$_2$ molecules, we recommend the use of these new rates in non-LTE models of CH$_3$CN excitation.

A systematic comparison of the excitation rate coefficients for CH$_3$CN-He and CH$_3$NC-He shows that the rates differ by up to a factor of 3. These isomerism effects in the collisional excitation show that separate calculations must be performed for each isomer as the rate coefficients cannot be assumed to be equal. These differences also let us predict that the excitation of these molecules in cold molecular clouds will be different, and we must use the appropriate rate coefficients of each isomer.
To confirm this, we carried out a radiative transfer simulation for both isomers using scaled rate coefficients of CH$_3$CN/NC. As predicted, the comparison of both brightness and excitation temperatures demonstrates that the excitation is different from one isomer to another and confirm again the need for separate analysis for both isomers.

In this paper, the collisional excitation of CH$_3$CN and CH$_3$NC with He atoms was carried out. He is often considered as a proxy for the spherical $para-$H$_2$($j$=0), especially for large molecules (more than 6 atoms) due to the larger complexity of the interaction and dynamics with H$_2$ (and the required computational resources needed to perform the calculations).
Although this approximation is useful, it is often found that H$_2$ behaves differently than He in excitation studies, in particular for collisions involving $ortho-$H$_2$. This deserves its own studies and will be the subject of future work.

\section*{Acknowledgements}
J.L. acknowledges support from KU Leuven through Grant No. 19-00313.












\bsp	
\label{lastpage}
\end{document}